\documentclass[longauth]{aa}

\usepackage[utf8]{inputenc}
\usepackage{graphicx}
\usepackage{amssymb}
\usepackage{amsmath}
\usepackage{float}
\usepackage{natbib}
\bibpunct{(}{)}{;}{a}{}{,} 
\usepackage{array}
\usepackage{hyperref}
\usepackage{subfig}
\usepackage[varg]{txfonts}
\usepackage{multirow}
\usepackage{siunitx}
\usepackage{upgreek}
\usepackage{textgreek}
\usepackage{ulem}
\usepackage{titlesec}
\usepackage{tablefootnote}
\usepackage{lscape}
\usepackage{xcolor}

\hypersetup{colorlinks=true,citecolor=blue}

\titleformat{\paragraph}
{\normalfont\normalsize\bfseries}{\theparagraph}{1em}{}
\titlespacing*{\paragraph}
{0pt}{3.25ex plus 1ex minus .2ex}{1.5ex plus .2ex}

\def\app#1#2{%
  \mathrel{%
    \setbox0=\hbox{$#1\sim$}%
    \setbox2=\hbox{%
      \rlap{\hbox{$#1\propto$}}%
      \lower1.1\ht0\box0%
    }%
    \raise0.25\ht2\box2%
  }%
}

\DeclareTextSymbol{\degre}{T1}{6}
\DeclareTextSymbol{\degre}{OT1}{23}

\makeatletter
\g@addto@macro{\endtabular}{\rowfont{}}
\makeatother
\newcommand{\rowfonttype}{}
\newcommand{\rowfont}[1]{
\gdef\rowfonttype{#1}#1\ignorespaces%
}
\makeatother

\defcitealias{}{}          

\begin{document}
  \title{WASP-127b: A misaligned planet with a partly cloudy atmosphere and tenuous sodium signature seen by ESPRESSO\thanks{Based on Guaranteed Time Observations collected at the European Southern Observatory under ESO programme 1102.C-0744 by the ESPRESSO Consortium.}}

   \author{ R. Allart\inst{1,*} ,
   L. Pino\inst{2,3},
   C. Lovis\inst{1},
   S. G. Sousa\inst{4},
   N. Casasayas-Barris\inst{5,6},
   M. R. Zapatero Osorio\inst{7},
   M. Cretignier\inst{1},
   E. Palle\inst{5,6},
   F. Pepe\inst{1},
   S. Cristiani\inst{8,9},
   R. Rebolo\inst{5,6,10},
   N.C. Santos\inst{4,11},
   F. Borsa\inst{12},
   V. Bourrier\inst{1},
   O.D.S. Demangeon\inst{4,11},
   D. Ehrenreich\inst{1},
   B. Lavie\inst{1},
   J. Lillo-Box\inst{7},
   G. Micela\inst{13},
   M. Oshagh\inst{5,6},
   A. Sozzetti\inst{14},
   H. Tabernero\inst{4,11},
   V. Adibekyan\inst{4,11},
   C. Allende Prieto\inst{5,6},
   Y. Alibert\inst{15},
   M. Amate\inst{5,6},
   W. Benz\inst{15},
   F. Bouchy\inst{1},
   A. Cabral\inst{16,17},
   H. Dekker\inst{18},
   V. D’Odorico\inst{8,9},
   P. Di Marcantonio\inst{8},
   X. Dumusque\inst{1},
   P. Figueira\inst{19,4},
   R. Genova Santos\inst{5,6},
   J. I. Gonz\'alez Hern\'andez \inst{5,6},
   G. Lo Curto\inst{19},
   A. Manescau\inst{18},
   C.J.A.P. Martins\inst{4,22},
   D. Mégevand\inst{1},
   A. Mehner\inst{19},
   P. Molaro\inst{8,20},
   N. J. Nunes\inst{16,17},
   E. Poretti\inst{12,21}
   M. Riva\inst{14},
   A. Suárez Mascareño\inst{5,6},
   S. Udry\inst{1}
   and F. Zerbi\inst{12}}
   
   \institute{\inst{1} Observatoire astronomique de l'Universit\'e de Gen\`eve, Universit\'e de Gen\`eve, 51 chemin des Maillettes, CH-1290 Versoix, Switzerland\\
              		* \email{romain.allart@unige.ch}\\
              		\inst{2} Anton Pannekoek Institute for Astronomy, University of Amsterdam Science Park 904 1098 XH Amsterdam, The Netherlands\\
              		\inst{3} INAF-Osservatorio Astrofisico di Arcetri Largo Enrico Fermi 5 I-50125 Firenze, Italy\\
              		\inst{4} Instituto de Astrofísica e Ciências do Espaço, Universidade do Porto, CAUP, Rua das Estrelas, PT4150-762 Porto, Portugal\\
              		\inst{5} Instituto de Astrofísica de Canarias, Vía Láctea s/n, 38205 La Laguna, Tenerife, Spain\\
              		\inst{6} Departamento de Astrofísica, Universidad de La Laguna, Spain\\
              		\inst{7} Centro de Astrobiología (CSIC-INTA), ESAC, Camino bajo del castillo s/n, 28692 Villanueva de la Cañada, Madrid, Spain\\
              		\inst{8} INAF – Osservatorio Astronomico di Trieste, via Tiepolo 11, 34143 Trieste, Italy\\
              		\inst{9} Institute for Fundamental Physics of the Universe, IFPU, Via Beirut 2, 34151 Grignano, Trieste, Italy\\
              		\inst{10} Consejo Superior de Investigaciones Científicas, E-28006 Madrid, Spain\\
              		\inst{11} Departamento de Física e Astronomia, Faculdade de Ciências, Universidade do Porto, Rua do Campo Alegre 687, PT4169-007 Porto, Portugal \\    
              		\inst{12} INAF – Osservatorio Astronomico di Brera, Via E. Bianchi 46, 23807 Merate (LC), Italy\\   
              		\inst{13} INAF - Osservatorio Astronomico di Palermo , Piazza del Parlamento 1 , 90134 Palermo.\\
              		\inst{14} INAF – Osservatorio Astrofisico di Torino, Via Osservatorio 20, I-10025 Pino Torinese, Italy\\
              		\inst{15} Universitat Bern, Physikalisches Institut, Siedlerstrasse 5, 3012 Bern, Switzerland\\
              		\inst{16} Faculdade de Ciênçias da Universidade de Lisboa (Departamento de Física), Edificio C8, 1749-016 Lisboa, Portugal\\
              		\inst{17} Instituto de Astrofísica e Ciênçias do Espaço, Universidade de Lisboa, Edificio C8, 1749-016 Lisboa, Portugal\\
              		\inst{18} European Southern Observatory, Karl-Schwarzschild-Strasse 2, 85748 Garching b. Munchen, Germany\\
              		\inst{19} ESO, European Southern Observatory, Alonso de Cordova 3107, Vitacura, Santiago\\
              		\inst{20} Institute for Fundamental Physics of the Universe, IFPU, Via Beirut 2, 34151 Grignano, Trieste, Italy\\	
              		\inst{21} Fundación G. Galilei - INAF (TNG), Rambla J. A. Fernández Pérez 7, E-38712 Breña Baja (La Palma), Spain\\
              		\inst{22} Centro de Astrof\'{\i}sica da Universidade do Porto, Rua das Estrelas, 4150-762 Porto, Portugal \\
             }

   \date{Received January 1, 2015; accepted January 1, 2015}


  \abstract
    {The study of exoplanet atmospheres is essential to understand the formation, evolution and composition of exoplanets. The transmission spectroscopy technique is playing a significant role in this domain. In particular, the combination of state-of-the-art spectrographs at low- and high-spectral resolution is key to our understanding of atmospheric structure and composition.}
   {Two transits of the close-in sub Saturn-mass planet, WASP-127b, have been observed with ESPRESSO in the frame of the Guaranteed Time Observations Consortium. Transit observations allow us to study simultaneously the system architecture and the exoplanet atmosphere.}
    {We used the Reloaded Rossiter-McLaughlin technique to measure the projected obliquity $\lambda$ and the projected rotational velocity $\mathrm{v_{eq}\cdot sin(i_*)}$. We extracted the high-resolution transmission spectrum of the planet to study atomic lines. We also proposed a new cross-correlation framework to search for molecular species and we applied it to water vapor.}
   {The planet is orbiting its slowly rotating host star ($\mathrm{v_{eq}\cdot sin(i_*)}$\,=\,0.53$^{+0.07}_{-0.05}$\,km$\cdot$s$^{-1}$) on a retrograde misaligned orbit ($\lambda$\,=\,-\,128.41$^{+5.60}_{-5.46}$\,$^{\circ}$). We detected the sodium line core at the 9-$\sigma$ confidence level with an excess absorption of 0.34\,$\pm$\,0.04\,\%, a blueshift of 2.74\,$\pm$\,0.79\,km$\cdot$s$^{-1}$ and a FWHM of 15.18\,$\pm$\,1.75\,km$\cdot$s$^{-1}$. However, we did not detect the presence of other atomic species but set upper-limits of only few scale heights. Finally, we put a 3-$\sigma$ upper limit, to the average depth of the 1600 strongest water lines at equilibrium temperature in the visible band, of 38\,ppm. This constrains the cloud-deck pressure between 0.3 and 0.5\,mbar by combining our data with low-resolution data in the near-infrared and models computed for this planet.}
 {WASP-127b, with an age of about 10\,Gyr, is an unexpected exoplanet by its orbital architecture but also by the small extension of its sodium atmosphere ($\sim$7 scale heights).  ESPRESSO allows us to take a step forward in the detection of weak signals, thus bringing strong constraints on the presence of clouds in exoplanet atmospheres. The framework proposed in this work can be applied to search for molecular species and study cloud-decks in other exoplanets.}

   \titlerunning{WASP-127b seen by ESPRESSO}
   \authorrunning{R. Allart, L. Pino, C. Lovis et al. }
   \maketitle
%
\section{Introduction}
During the past few years, exoplanet atmospheric studies have tremendously grown. Detections of atomic and molecular species are reported from the ultraviolet (UV) to the infrared at low- and high-resolution for Earth-mass to Jupiter-mass planets and from temperate to ultra-hot worlds. This diversity of observational studies relies first on precise radii and masses gathered by extensive photometric (e.g., WASP, Kepler and TESS \citep{pollacco_wasp_2006,borucki_kepler_2003,ricker_transiting_2014} and radial velocities surveys (Coralie, HARPS, HARPS-N, ESPRESSO \citep{queloz_coralie_2000,mayor_setting_2003,cosentino_harps-n_2012,pepe_espressovlt_2020}. Second, it relies on a diversity of instruments able to characterize exoplanet atmospheres. Among the most efficient ones, there are space telescopes and ground-based high-resolution spectrographs. The \textit{Hubble Space Telescope} (\textit{HST}) was the spearhead for several years, with multiple detections of water in the near-infrared (NIR), atomic species in the visible, and a diversity of clouds and hazes \citep{kreidberg_clouds_2014,sing_continuum_2016}. \\
Nowadays, ground-based high-resolution spectrographs are starting to produce groundbreaking results that inform us on the temperature structure and dynamics of exoplanet atmospheres \citep[e.g.][]{snellen_orbital_2010,wyttenbach_spectrally_2015,louden_spatially_2015,brogi_rotation_2016,allart_spectrally_2018,allart_high-resolution_2019,seidel_hot_2019,casasayas-barris_atmospheric_2019,borsa_gaps_2019,ehrenreich_nightside_2020}. The advantage of the high-resolution technique is that it resolves absorption lines, allowing to remove systematics (e.g., stellar, telluric, instrumental contamination), to distinguish between atmospheric origin or Rossiter-McLaughlin effect \citep{rossiter_detection_1924,mclaughlin_results_1924,casasayas-barris_is_2020}, and also to derive dynamical properties of the atmosphere. However, this technique suffers heavily from the Earth's atmosphere contamination, which manifests as telluric absorption lines at specific wavelength that need to be corrected. Moreover, the Earth atmosphere introduces random flux variation that are uncalibrated. Therefore, the high-resolution technique prevents any absolute absorption measurement. Instead, it measures only excess absorption from the local continuum. Fortunately, such absolute measurements can be obtained from space-based low-resolution observations, and thus combining low- and high-resolution is key to have a complete view for each exoplanet studied but also a global picture of exoplanet atmospheres.

\section{A reference system for atmospheric studies: WASP-127b}\label{system WASP-127}
WASP-127b \citep{lam_dense_2017} orbits a bright (V=10.17) G5-type star with a period of 4.18 days. The host star is at the end of its main-sequence phase with an age of 9.7\,$\pm$\,1.0\,Gyr and radius of 1.30\,$\pm$\,0.04\,$R_{\odot}$ and has entered in the sub-giant phase. Therefore, the planet might live for the second time an inflation process \citep{lopez_re-inflated_2016}, leading to its large radius (1.31\,$\pm$\,0.03\,$R_{J}$). Other processes could also explain this puffiness, such as tidal heating \citep{bodenheimer_tidal_2001,bodenheimer_radii_2003}, ohmic heating \citep{batygin_evolution_2011}, or migration processes through Kozai effect \citep{kozai_secular_1962,fabrycky_shrinking_2007,bourrier_orbital_2018}. WASP-127b is a lukewarm ($\sim$600 times Earth irradiation) Saturn (0.165$\pm$0.025\,M$_J$) at the right mass border of the evaporation desert \citep{lecavelier_des_etangs_diagram_2007} with half the mass of Saturn but a radius larger than Jupiter and is thus a very low-density planet ($\sim$0.09\,g$\cdot$cm$^{-3}$). Assuming its mean molecular weight is 2.3\,$\mathrm{g\cdot mol^{-1}}$ (hydrogen-helium composition), its atmospheric scale height is about 2100\,km. In this case, the signal in transmission for one scale height is around 420\,ppm, making this planet, between those of its class, potentially one of the best to study exo-atmospheres through transmission spectroscopy. \\
The stellar and planetary parameters used for this study are given in Table\,\ref{parameterW127}. We derived the stellar mass, radius and age using the Padova stellar model isochrones (see.\cite{da_silva_basic_2006} and \cite{bressan_parsec_2012} - \url{http://stev.oapd.inaf.it/cgi-bin/param_1.3}), with the spectroscopic stellar parameters derived in this work together with the very precise \emph{Gaia} DR2 parallax (6.2409 +-0.0468 - \cite{gaia_collaboration_vizier_2018}). We derive $T_{\mathrm{eff}}$, $\log g$, microturbulence and [Fe/H] with their uncertainties using ARES+MOOG, following the methodology described in \citet[][]{sousa_ares_2014,santos_sweet-cat_2013}. We make use of equivalent widths (EW) of iron lines measured using the ARES code\footnote{The last version of ARES code (ARES v2) can be downloaded at http://www.astro.up.pt/$\sim$sousasag/ares} \citep{sousa_new_2007, sousa_ares_2015} on the combined ESPRESSO spectrum using the list of lines presented in \citet[][]{sousa_spectroscopic_2008}. A minimization process assuming ionization and excitation equilibrium is used to find convergency for the best set of spectroscopic parameters. Here we make use of a grid of Kurucz model atmospheres \citep{kurucz_synthe_1993} and the radiative transfer code MOOG \citep{sneden_carbon_1973}.\\
Due to its recent discovery, only a few atmospheric studies have been reported so far. Visible low-resolution ground-based observations \citep{palle_feature-rich_2017,chen_gtc_2018} indicate a rich atmosphere with the presence of Na, K, Li, and hazes. The sodium detection was confirmed at high-resolution with HARPS \citep{zak_high-resolution_2019}, but the same data have been reanalyzed by \cite{seidel_hot_2020} showing a flat Na spectrum. More recently, \textit{HST}/STIS and \textit{HST}/WFC3 data have been combined \citep{spake_abundance_2020, skaf_ares_2020}. \cite{spake_abundance_2020} has higher precision than \cite{palle_feature-rich_2017,chen_gtc_2018} and report a non-detection of Li, a hint of K, a detection of Na (5-$\sigma$), the presence of hazes and clouds and the presence of molecular absorptions (H$_2$O in the J-band, and possibly CO$_2$ at 4.5 microns). Moreover, the water detection reported is the strongest one observed in an exoplanet with a mean amplitude around 800\,ppm while previous detections around Hot-Jupiters were around 200\,ppm \citep{deming_infrared_2013,mccullough_water_2014}. Such planets with high amplitude water signature in the J-band are well placed to be studied from the ground with near-infrared spectrographs, allowing to analyze the impact of hazes and clouds by measuring the water content at different wavelengths \citep{de_kok_identifying_2014,pino_diagnosing_2018,alonso-floriano_multiple_2019,sanchez-lopez_water_2019}.\\
ESPRESSO observations of WASP-127b will not only help resolve the debate on the presence of atomic species but also to search for water vapor and thus constrain the properties of the planetary cloud-deck.

In Sect\,\ref{Sec_obs} we describe the ESPRESSO data used in this paper. In Sect.\,\ref{Sec_RM}, we analyse the Rossiter-McLaughlin effect. In Sect.\,\ref{Sec_method}, we describe the methods used to search for atomic species and water vapor. Sect.\,\ref{Sec_analysis} shows the results which are then discussed in Sect.\,\ref{Sec_interp}. We conclude in Sect.\,\ref{Sec_concl}.

\begin{table*}[h]
\centering
\caption{\footnotesize Adopted physical and orbital parameters for the WASP-127 system.}\label{parameterW127}
\begin{tabular}{lccc}
\hline
Parameter & Symbol & Value & Reference\\
\hline
\multicolumn{4}{c}{ \textit{Stellar parameters}} \\
\hline
Stellar mass & $M_{*}$ & 0.960\,$\pm$\,0.023\,$M_{\odot}$ & This work  \\
Stellar radius & $R_{*}$ & 1.303\,$\pm$\,0.037\,$R_{\odot}$ & This work  \\
Effective temperature & T$_{eff}$ & 5842\,$\pm$\,14\,K & This work\\
Metallicity & [Fe/H] & -0.19\,$\pm$\,0.01 & This work \\
surface gravity & log(g) & 4.23\,$\pm$\,0.02\,cgs & This work \\
Age & & 9.656\,$\pm$\,1.002\,Gyr & This work\\
\hline
\multicolumn{4}{c}{\textit{Planet parameters}} \\
\hline
Planet mass & $M_{p}$ & 0.165\,$\pm$\,0.021\,$M_{J}$ & \cite{seidel_hot_2020}  \\
Planet radius & $R_{p}$ & 1.311\,$\pm$\,0.025\,$R_{J}$ & \cite{seidel_hot_2020} \\
White-light radius ratio & $R_{p}/R_{*}$ & 0.10103441\,$\pm$0.00047197\,  & \cite{seidel_hot_2020} \\
Epoch of transit & $T_{0}$ & 2456776.621238\,$\pm$\,0.00023181\,BJD & \cite{seidel_hot_2020}  \\
Duration of transit & $T_{14}$ & 0.18137185\,$\pm$\,0.00035381\,d & \cite{seidel_hot_2020}  \\
Orbital period & $P$ & 4.17806203\,$\pm$\,0.00000088\,d & \cite{seidel_hot_2020}\\
Systemic velocity & $\gamma$ & -9295.46\,$\pm$\,1.4 \,m\,s$^{-1}$ & This work \\
Semi-amplitude & K$_*$ & 21.51\,$\pm$\,2.78\,m\,s$^{-1}$ & \cite{seidel_hot_2020}  \\
Eccentricity & $e$ & 0.0 & \cite{seidel_hot_2020}  \\
Argument of the periastron &$\omega$ & 0.0 & \cite{seidel_hot_2020}  \\
Scaled semi-major axis & $a$/$R_{*}$ & 7.808\,$\pm$\,0.109 & \cite{seidel_hot_2020}  \\
Inclination & $i_p$ & 87.84\,$\pm$\,0.36\,$^{\circ}$ & \cite{seidel_hot_2020} \\
Impact parameter & b & 0.29\,$\pm$\,0.045 & \cite{seidel_hot_2020} \\
Limb-darkening & $u_1$ & 0.422 & \cite{seidel_hot_2020} \\
Limb-darkening  & $u_2$ &  0.214 & \cite{seidel_hot_2020} \\
\hline
\end{tabular}
\end{table*}


\section{ESPRESSO observations}\label{Sec_obs}
We observed two transits of WASP-127b (2019-02-24 and 2019-03-17) with the \textit{Echelle SPectrograph for Rocky Exoplanets and Stable Spectroscopic Observations} (ESPRESSO, \cite{pepe_espressovlt_2020}). ESPRESSO is a fiber-fed, ultra-stabilized high-resolution echelle spectrograph installed at Paranal. It can collect the light for any or all 8\,m Unit Telescope (UT) of the Very Large Telescope (VLT). The observations have been obtained in the frame of the GTO consortium (Prog: 1102.C-0744, PI: F. Pepe) on UT2 in the HR21 mode ($\mathcal{R}\sim$140000).  The fiber B was used to monitor the sky. Due to instrumental limitations of the atmospheric dispersion compensator (ADC), exposures above airmass 2.2 have been removed (the three last exposures of the first night). The first night has a globally higher Signal-to-Noise Ratio (S/N), but it is more impacted by water column content, which impacts the search for water vapor (Sect\,\ref{Sec_method} and \ref{Sec_analysis}). We summarized the night conditions in Table \ref{observation W127}.\\
The data have been reduced using the version 2.0.0 Data Reduction Software (DRS) pipeline. As moon contamination is present on both nights, we used the \texttt{CCF\_SKYSUB\_A}\footnote{\texttt{CCF} stands for cross correlation function used to measure the radial velocity.} and \texttt{S1D\_SKYSUB\_A}\footnote{\texttt{S1D} stands for one dimentional spectrum after order merging.} products, which include the subtraction of the sky contribution simultaneously recorded of fiber B, at the cost of a marginally lower S/N.

\begin{table}[h]
\centering
\caption{Observational log of the two nights.}
\begin{tabular}{lcccc}
\hline
Observing nights & 2019-02-24 & 2019-03-17\\
\hline
Total spectra & 71 (74\tablefootnote{This is the total exposures number including those below airmass 2.2}) & 74 \\
In-transit & 43 & 39 \\
Out-of-transit\tablefootnote{Number of exposures before and after transit} & 19-9 & 13-18 \\
 t$_{exp}$ [s] & 300 & 300 \\
 Airmass & 1.07 - 2.17 & 1.07 -1.91 \\
 Seeing & 0.28 - 2.05 & 0.45 - 1.13\\
 S/N@550nm & 32 - 88 & 43 - 77 \\
 PWV [mm] & 5.8 - 7.9 & 3.0 - 3.7\\
\hline
\end{tabular}
\label{observation W127}
\end{table}


\section{Reloaded Rossiter-McLaughlin analysis}\label{Sec_RM}
\begin{figure*}[h]
\resizebox{\hsize}{!}{\includegraphics[height=\textheight]{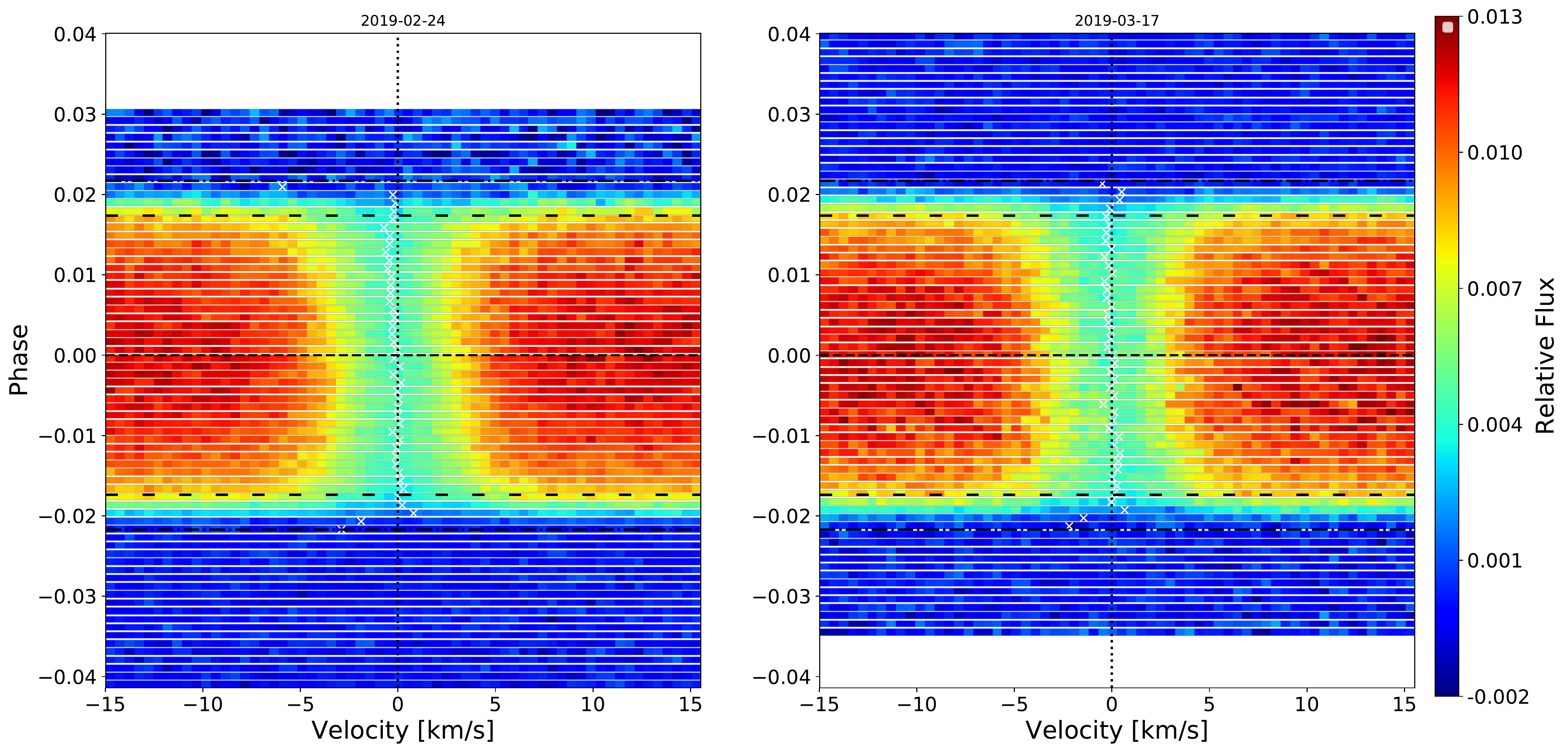}}
\caption[WASP-127b_local_CCFs_map]{Local CCF map for both nights. Each row corresponds to a local CCF expressed in velocity in the star rest frame at a given time. The two dashed dotted horizontal black lines represent the contact points $t_1$ and $t_4$, the dashed horizontal black lines are $t_2$ and $t_3$, while the vertical dashed black line is at null velocity. The white cross are the local velocities derived by the gaussian fit. The RM effect is clearly detected in both nights as the green trace.}
\label{WASP-127b_local_CCFs_map}
\end{figure*}

\begin{figure*}[h]
\resizebox{\hsize}{!}{\includegraphics[height=\textheight]{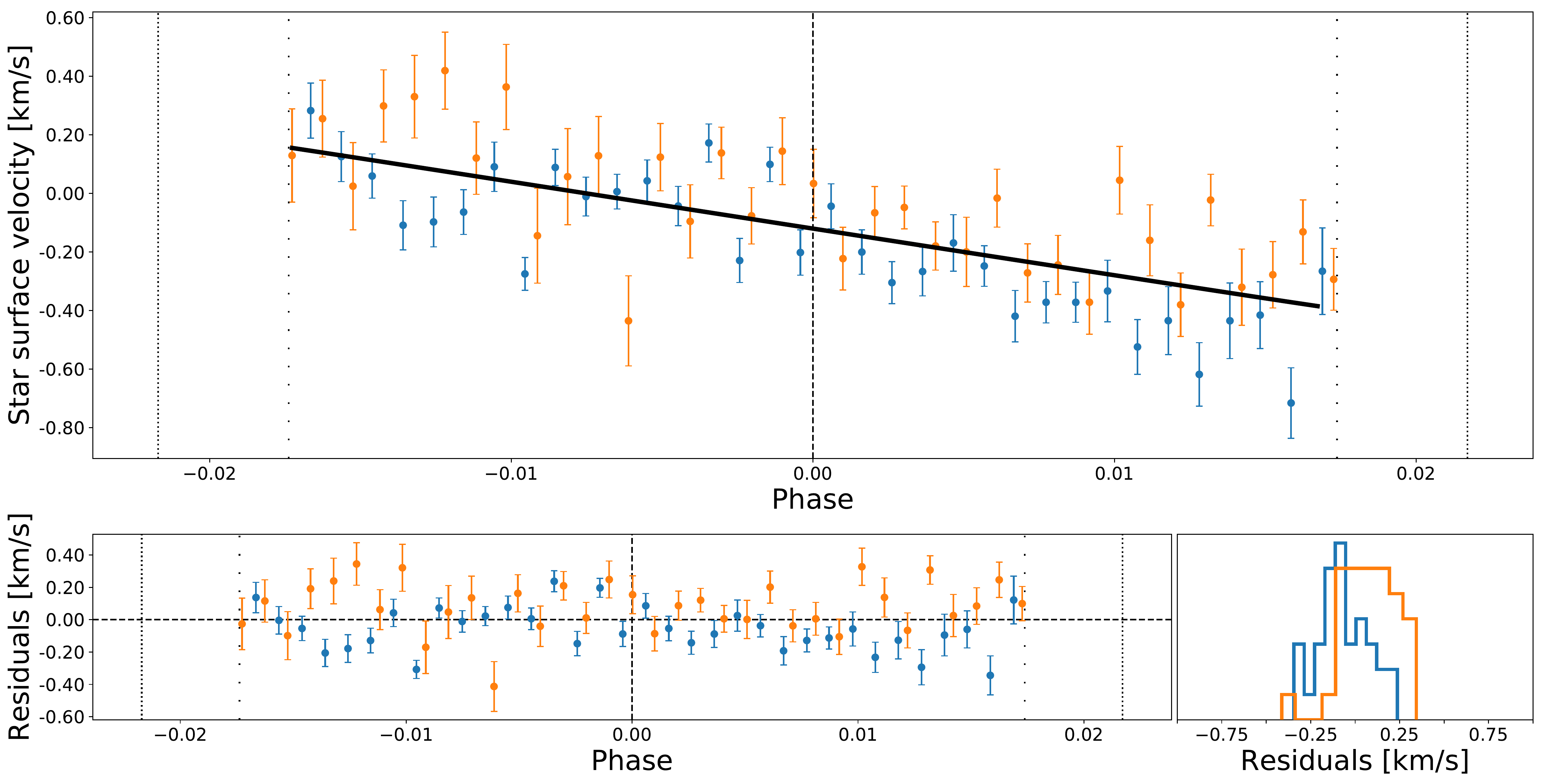}}
\caption[WASP-127b_best_model_nested_contrast]{\textit{Top}: Extracted local velocities for both nights (respectively blue and orange for the 2019-02-24 and 2019-03-17) with the best-fit model in black. The vertical dotted lines represent the contact points $t_1$ and $t_4$, the vertical loosely dotted lines represent the contact points $t_2$ and $t_3$ and the vertical dashed line represents the mid-transit. \textit{Bottom left}: The local velocity residuals against the best-fit model for both nights. The vertical lines are similar as the top panel and the horizontal dashed line represents null velocity. \textit{Bottom right}: Distribution of the residuals in the bottom left panel.}
\label{WASP-127b_best_model_nested_contrast}
\end{figure*}

\begin{figure}[h]
\resizebox{\hsize}{!}{\includegraphics[height=\textheight]{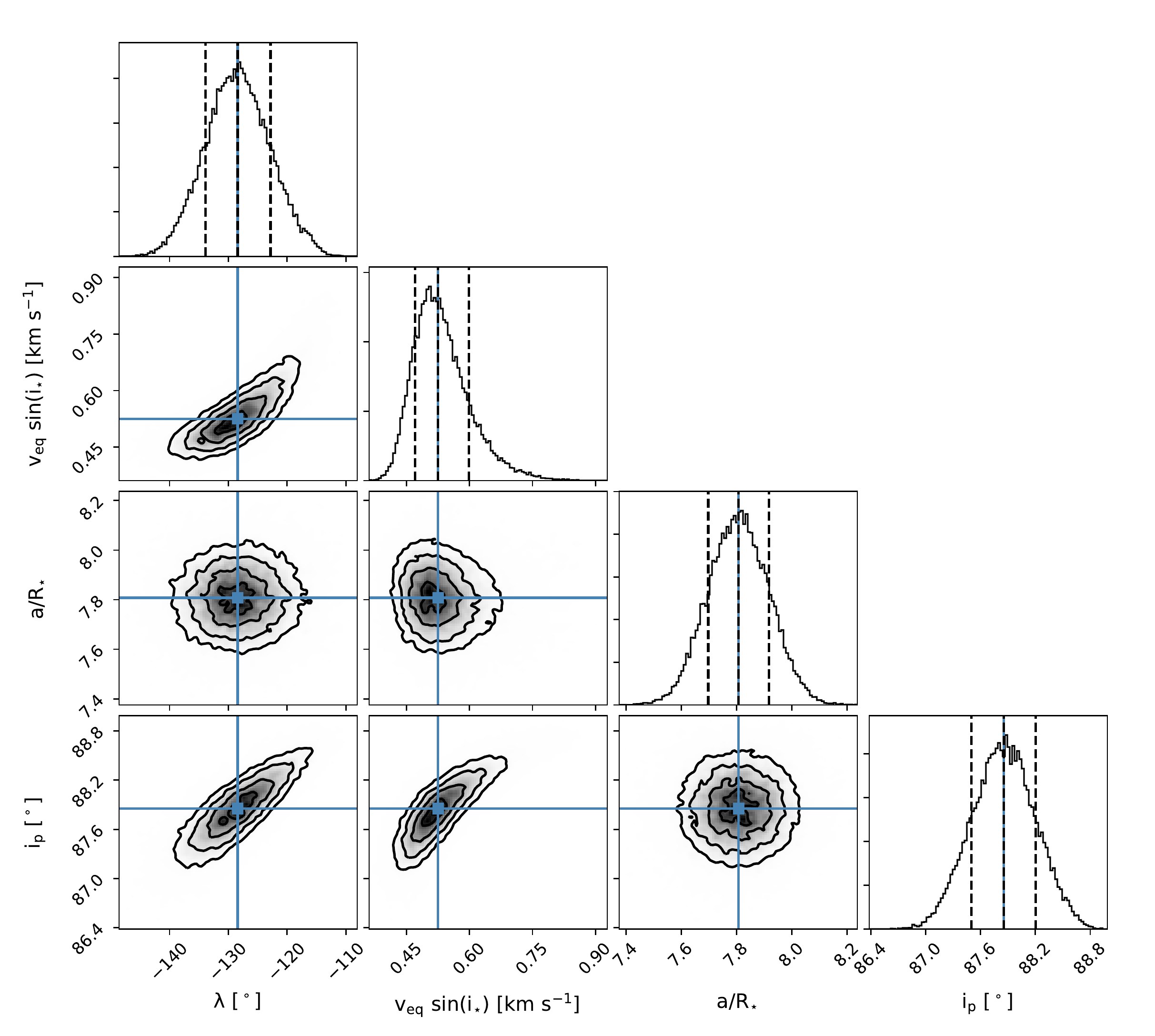}}
\caption[corner_rigid_body_notitle_median]{Posterior distribution of the best-fit Rossiter-McLaughlin model done with a Nested sampling and 10000 living points.}
\label{corner_rigid_body_and_b_notitle_median}
\end{figure}

\begin{figure}[h]
\resizebox{\hsize}{!}{\includegraphics[height=\textheight]{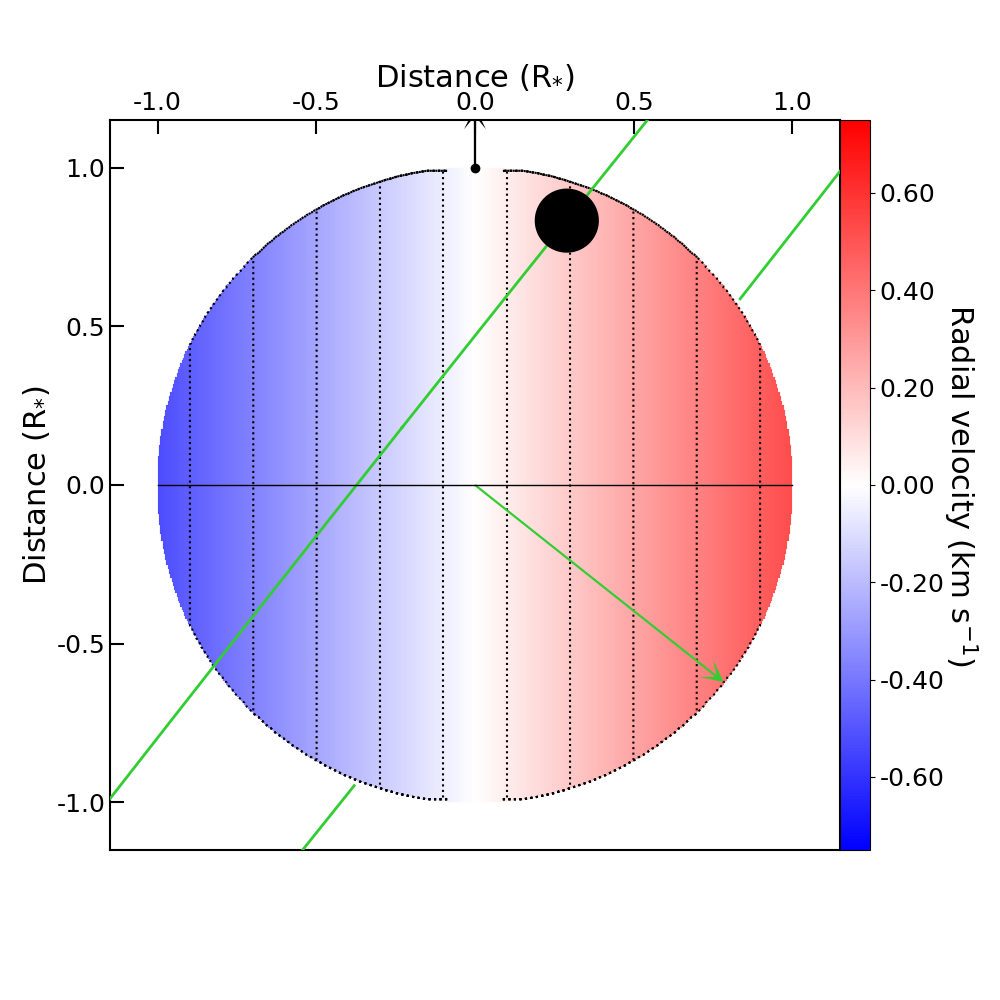}}
\caption[WASP-127b_st_disk_ross_rigid_body]{View of the star planet system at the ingress transit. The stellar disk velocity has a color gradient from blue to red for negative to positive stellar surface velocity. The stellar rotation axis is shown as the black top arrow. The planet is represented by the black disk and occults first the red part of the star and then the blue part, following its misaligned orbit shown in green. The green arrow represents the orbital axis.}
\label{WASP-127b_st_disk_ross_rigid_body}
\end{figure}

We analyzed the RM effect with the Reloaded RM technique \citep{cegla_rossiter-mclaughlin_2016,bourrier_orbital_2018,ehrenreich_nightside_2020}. It consists in extracting the stellar surface radial velocity behind the planet at each exposure. The disk-integrated CCFs are Doppler shifted to the stellar rest frame using the orbital solution (see Table\,\ref{parameterW127}). We flux-calibrated the CCFs by first normalizing each CCF by its continuum level derived from a Gaussian fit and second by scaling the normalized CCFs to the planetary transit depth at their respective time. To do so, we used the \texttt{Batman} python package \citep{kreidberg_batman:_2015} to compute a theoretical transit light curve based on the parameters given in Table\,\ref{parameterW127}. The rescaled CCFs were then Doppler shifted by the velocity of the master out-of-transit CCF making the results independent of velocity offsets and then interpolated on a same velocity grid. We extracted the occulted stellar surface CCFs (hereafter local CCFs) by subtracting each scaled in-transit CCF from the master out-of-transit CCF (Fig.\,\ref{WASP-127b_local_CCFs_map}). A Gaussian fit is applied to each local CCFs to derive the projected velocity of the stellar surface behind the planet. Local CCFs during ingress and egress have been excluded as they have lower S/N and are more sensitive to the limb-darkening law parameters. \\
The local velocities (shown in Fig.\,\ref{WASP-127b_best_model_nested_contrast}) roughly follow a straight line, indicative of solid-body rotation and are first redshifted, and later blueshifted, which indicates a retrograde orbit. We fitted the local velocities with the model of stellar rotation described in \cite{cegla_rossiter-mclaughlin_2016,bourrier_refined_2017,bourrier_hot_2020,ehrenreich_nightside_2020} to constrain the projected rotational velocity, $\mathrm{v_{eq}\cdot sin(i_*)}$, the projected obliquity, $\lambda$, and the impact parameter, which we expressed as $\mathrm{a/R_*\cdot cos(i_p)}$. We perform a Bayesian inference on the model's parameters using the nested sampling algorithm \citep{skilling_calibration_2006} via the python package pymultinest \citep{buchner_x-ray_2014,feroz_multimodal_2008,feroz_multinest_2009,feroz_exploring_2013}, which is widely used throughout exoplanet research \citep[e.g.][]{lavie_helios-retrieval_2017, peretti_orbital_2019, seidel_wind_2020, ehrenreich_nightside_2020}. We use a gaussian likelihood ($\mathcal{L}$) and the priors are summarized in table\,\ref{Prior_MCMC}. From spectroscopic measurements, the star is known to be a slow rotator, we thus set a tight uniform prior on $\mathrm{v_{eq}\cdot sin(i_*)}$. The Gaussian priors on $\mathrm{a/R_*}$ and $\mathrm{i_p}$ are the posteriors of simultaneous Euler photometry described in \cite{seidel_hot_2020}.\\
The nested sampling algorithm is run with 10000 live points to ensure a thorough exploration of the parameter space. The best-fit model is reported in Fig.\,\ref{WASP-127b_best_model_nested_contrast} and is obtained with a log($\mathcal{L}$) value of -19.12 for $\lambda$\,=\,-128.41$^{+5.60}_{-5.46}$\,$^{\circ}$, $\mathrm{v_{eq}\cdot sin(i_*)}$\,=\,0.53$^{+0.07}_{-0.05}$\,km$\cdot$s$^{-1}$, $\mathrm{a/R_*}$\,=\,7.81$^{+0.11}_{-0.11}$ and $\mathrm{i_p}$\,=\,87.85$^{+0.35}_{-0.35}$\,$^{\circ}$. The 1-$\sigma$ uncertainties are obtained from the posterior distributions shown in Fig.\,\ref{corner_rigid_body_and_b_notitle_median}. The posterior distributions on $\mathrm{a/R_*}$ and $\mathrm{i_p}$ are completely compatible with their prior distributions and the parameters are uncorrelated meaning that they are not constrained by the fit. There is no correlation between $\lambda$ and $\mathrm{v_{eq}\cdot sin(i_*)}$ as expected since the impact parameter is different from 0 \citep{burrows_near-infrared_2000,bourrier_hot_2020,ehrenreich_nightside_2020}. However, $\lambda$ and $\mathrm{v_{eq}\cdot sin(i_*)}$ show some correlation with $\mathrm{i_p}$. This is a known degeneracy \citep{albrecht_obliquities_2012,brown_rossiter-mclaughlin_2017,bourrier_hot_2020}.\\
Our best-fit model shows that the old star WASP-127 is a slow rotator while its planet has a misaligned retrograde orbit (a view of the system is represented in Fig.\,\ref{WASP-127b_st_disk_ross_rigid_body}). We also note that the WASP-127 system does not fit in the known dichotomy of hot exoplanets. \cite{winn_hot_2010,albrecht_obliquities_2012} have reported that stars with T$_{eff}$ below 6250\,K have aligned systems which is not the case of WASP-127b (T$_{eff}$\,=\,5842\,$\pm$\,14\,K). One possible scenario is that WASP-127b remained trapped in a Kozai resonance with an outer companion for billions of years, and only recently migrated close to its star (see e.g. the case of GJ436b, \cite{bourrier_orbital_2018}). An in depth analysis of the system dynamic is beyond the scope of the present study.

\begin{table}[h]
\centering
\small
\caption{Priors on the stellar rotation model parameters.}
\begin{tabular}{lccc}
\hline
Parameter & Prior type & Prior values & Posterior\\
\hline
$\lambda$ & Uniform & -180 - 180$^{\circ}$ & -128.41$^{+5.60}_{-5.46}$\,$^{\circ}$\\
$\mathrm{v_{eq}\cdot sin(i_*)}$ & Uniform & 0 - 3\,km$\cdot$s$^{-1}$ & 0.53$^{+0.07}_{-0.05}$\,km$\cdot$s$^{-1}$ \\
$\mathrm{a/R_*}$ & Gaussian & $\mu$\,=\,7.81; $\sigma$\,=\,0.11 & 7.81$^{+0.11}_{-0.11}$ \\
$\mathrm{i_p}$ & Gaussian & $\mu$\,=\,87.84; $\sigma$\,=\,0.36  & 87.85$^{+0.35}_{-0.35}$\,$^{\circ}$\\
\hline
\end{tabular}
\label{Prior_MCMC}
\end{table}

\section{Extraction of planetary atmospheric signal}\label{Sec_method}
In this section we describe the reduction steps used to derive the final products: the transmission spectrum and the planetary CCFs. We first correct for telluric contamination (\ref{Sub_sec_tell}) then we normalized the spectra (\ref{Sub_sec_norm}) before deriving the transmission spectrum (\ref{Sub_sec_TS}) and the CCFs to search for water vapor (\ref{Sub_sec_CCF}).
\subsection{Telluric correction}\label{Sub_sec_tell}
Ground-based spectra are impacted by the Earth's atmosphere in two different ways. First, the absolute flux is lost, so calibration is needed to recover it (this is why ground-based transmission spectroscopy is a differential process where in-transit observations are compared to out-of-transit observations). Second, telluric band features absorb the stellar flux at precise wavelengths depending on the species (e.g., H$_2$O, O$_2$, CH$_4$, CO$_2$ or O$_3$). At high resolution, these bands are resolved as a forest of lines. Moreover, emission lines formed at higher altitudes in the Earth's atmosphere are also spectrally resolved, such as Na, O, or OH. This telluric contamination has a more significant impact at redder wavelengths but strong H$_2$O and O$_2$ bands are already present in the ESPRESSO wavelength range.\\

We used the version 1.5.1 of the \texttt{Molecfit} \citep{smette_molecfit:_2015,kausch_molecfit_2015} software following the method described in \citep{allart_search_2017}. We adjusted the fit region to the stronger telluric lines that are present in ESPRESSO. To investigate the water residuals, we computed a CCF with a binary telluric mask (see \ref{Sub_sec_CCF} and \ref{Sec_analysis}) on the transmission spectrum in the stellar rest frame (which across a night has only a constant shift with respect to the observer rest frame). Due to the higher water vapor content on the first night, a region of 1\,\AA-wide (exclusion of the pixels) around each of the 20 deepest telluric lines has been masked. It is likely that once the telluric lines reach a given depth (70-90\,\%), telluric models are missing some physics that prevents a perfect correction.\\

\subsection{Normalization}\label{Sub_sec_norm}
It is necessary to properly normalize the spectra to extract planetary information. Moreover, ESPRESSO spectra are affected by two interference patterns induced by coude train optics. They become apparent when spectra taken in different telescope positions are divided by each other, creating sinusoidal "wiggles"\footnote{A forthcoming paper will explain in details the properties of those wiggles}. The first set of wiggles have a period of 30\,\AA\, at 600\,nm and amplitude of $\sim$1\,\% (Tabernero et al. submitted, Borsa et al. accepted, Casasayas-Barris et al. submitted). In contrast, the second wiggles have a shorter period of 1\,\AA\, at 600\,nm and amplitude of $\sim$0.1\,\% (Casasayas-Barris et al. submitted). Our WASP-127b observations have too low S/N to be sensitive to the second wiggles while the first wiggles are detectable and need to be corrected. Contrarily to previous papers (Tabernero et al. submitted, Borsa et al. accepted, Casasayas-Barris et al. submitted), we decided to correct the wiggles directly on the stellar spectra and not on the transmission spectrum.\\
To do so, we are using the publicly available \texttt{RASSINE} code \citep{cretignier_rassine_2020}. \texttt{RASSINE} is a python code using an alpha shape strategy \citep{xu_modeling_2019} to normalize the spectrum, taking advantage of non-parametric models to fit the upper envelope of the merged 1d spectrum. The authors showed that alpha shape algorithms could provide more precise and more accurate continuum normalization than classical iterative fitting methods. The only hypothesis behind the algorithm is that the local maxima are probing the continuum of the spectrum. \\
To fit the wiggles with a periodicity of 30\,\AA\, the alpha shape needs to have a smaller radius in order to capture the modulation. Its value was selected to be about 4\,\AA\ here. Appendix\,\ref{AppendixE} shows the quality of this normalization.
\subsection{Transmission spectrum}\label{Sub_sec_TS}
To derive the transmission spectrum, we followed the methods described in \cite{wyttenbach_spectrally_2015,allart_search_2017}, which consist in Doppler shifting the normalized spectrum from the Earth barycentric rest frame to the stellar rest frame by taking into account the systemic velocity and the star radial velocity (from the Keplerian). Then we computed the master-out spectrum by averaging out-of-transit spectra. Individual transmission spectra are obtained by dividing each spectrum by the master-out spectrum. These individual transmission spectra can be used to study the origin of features and their evolution in time (see section\,\ref{Sub_sub_sec_Na}). They are then Doppler-shifted to the planet rest frame and averaged. For the atomic species, we applied a weighted average to properly take into account the low core flux of stellar lines (Fig.\,\ref{measured_CCF_Na_and_map}, left panel). Weights are set to one over the uncertainties square.

\subsection{Cross-correlation function - CCF}\label{Sub_sec_CCF}
The cross-correlation function is used to maximize the S/N by co-adding hundreds of lines together. In the exoplanets field, two different methods are often used to compute CCFs. The first one is mainly used to derive stellar radial velocities and consists of cross-correlating a spectrum with a binary mask \citep{baranne_elodie_1996,pepe_coralie_2002}. This technique is also used for atmospheric characterization \citep{allart_search_2017,pino_diagnosing_2018,pino_neutral_2020}. The second technique consists of cross-correlating a spectrum with a template or model, and it is mainly used for atmospheric characterization \citep{snellen_orbital_2010, birkby_detection_2013, brogi_rotation_2016, hoeijmakers_atomic_2018} but also for stellar radial velocities \citep{astudillo-defru_harps_2015,anglada-escude_terrestrial_2016}. We discuss more in detail the advantages and disadvantages of both methods in Appendix\,\ref{AppendixA}. In this paper, we focus on the search for water vapor by applying the binary mask methods to the water band between 7000 and 7500 \AA\ using the HITRAN database \citep{gordon_hitemp_2010} for the telluric mask (\ref{Sub_sec_tell} and \ref{Sec_analysis}) and HITEMP for exoplanetary masks \citep{rothman_hitemp_2010}. We discuss in Appendix\,\ref{AppendixB}, the impact of different line-lists in the visible and for our work.\\
To search for water vapor, we go further than what was previously done in \cite{allart_search_2017} by exploring in more detail the temperature and the number of lines parameters (as discussed in Appendix\,\ref{AppendixB}). The atmospheric temperature is explored from 100 to 3000\,K with a 100\,K step, and the number of lines is explored from 100 to 3000 with a 100-lines step. We produced CCFs for each set of temperatures and number of lines sampling the velocity space between -50 and 50\,km$\cdot$s$^{-1}$ with a step of to the pixel size, 0.5\,km$\cdot$s$^{-1}$. We then fitted a Gaussian on each CCF where the central velocity, the FWHM, and amplitudes were respectively initialized to 0\,km$\cdot$s$^{-1}$, 2.5\,km$\cdot$s$^{-1}$ and 100\,ppm. We restricted the FWHM to value above 2\,km$\cdot$s$^{-1}$ in the fit to at least sample one ESPRESSO line spread function (LSF). We reported on 2D maps (temperature vs number of lines): the detection level (amplitude divided by its uncertainty), the amplitude, the FWHM, and central velocity. We applied this method on the data in Sect.\,\ref{Sub_sec_Water} and in Appendix\,\ref{AppendixC}, where we injected a cloud-free model into the data to validate the procedure.


\section{Transmission spectroscopy analysis}\label{Sec_analysis}
In this section, we look at atomic lines (\ref{Sub_sec_Atomic}) that have previously been reported in different exoplanets such as \ion{Na}{i}, \ion{K}{i}, \ion{Li}{i}, and H-$\alpha$. We also search for water vapor in the subsection\,\ref{Sub_sec_Water}.

\subsection{Telluric contamination}
\begin{figure*}[h]
\includegraphics[width=\textwidth]{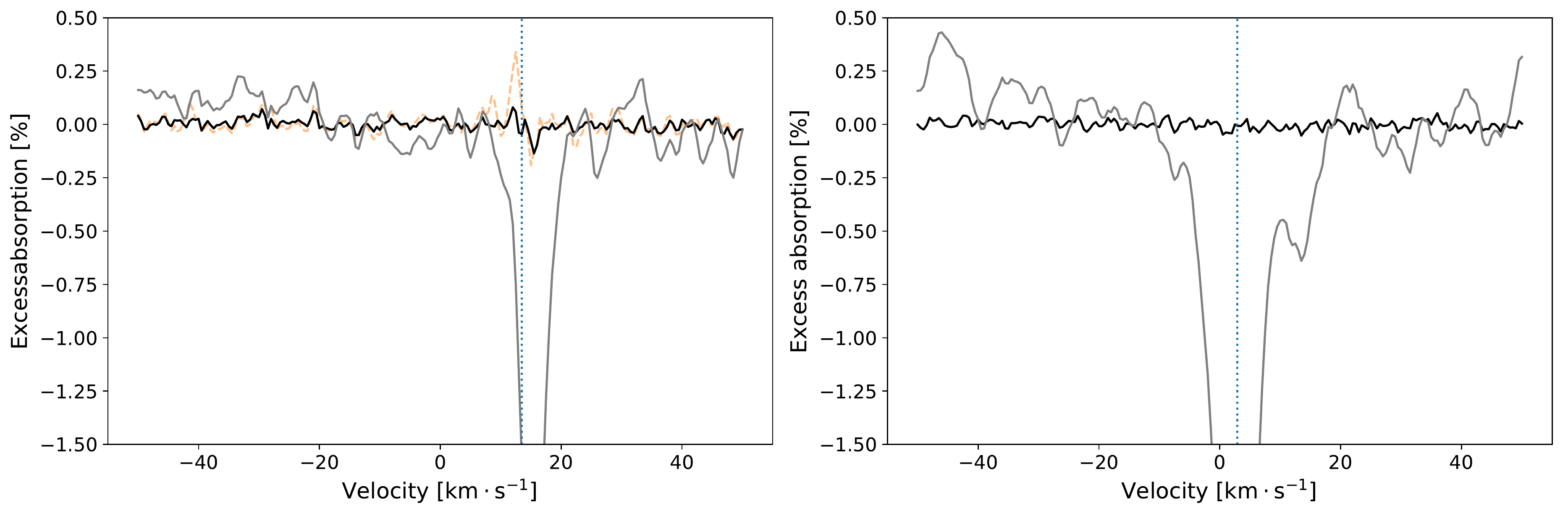}
\caption[Telluric_ccf]{CCFs of the transmission spectrum corrected from the telluric features (black) and non-corrected (gray) in the stellar rest frame using the
water vapor mask at 296 K for both nights (2019-02-24 and 2019-03-17 from left to right). For the 2019-02-24 night, the dashed orange CCF shows the impact of 20 deepest telluric lines if they are not masked. The blue dashed line shows the observer's radial velocity, where telluric residuals would be expected.}
\label{Telluric_ccf}
\end{figure*}

Fig.\,\ref{Telluric_ccf} shows the CCF of the transmission spectrum corrected from telluric contamination (black) for both nights obtained with the mask at 296\,K (ambient temperature) between 7000 and 7500\,\AA\ in the stellar rest frame. The mask is composed of the 150 strongest water telluric lines. If no telluric correction is applied, the corresponding CCFs (in grey) exhibit features respectively at 3\,\% and 5\,\%. Our telluric correction reduces this feature to the noise level (1-pixel dispersion of 250 and 189\,ppm). Nonetheless, the first night has still some telluric residuals around 0.2\,\% peak-to-valley amplitude over a few pixels despite the telluric masking applied (see Sect.\,\ref{Sub_sec_tell}). However, when we Doppler-shift to the planet rest frame, no residuals impact potential water signatures at null velocity.

\subsection{Atomic species}\label{Sub_sec_Atomic}
For the atomic species, the transmission spectrum was built on each night as the weighted average of the individual transmission spectra. This is done to avoid decreasing the final S/N due to the low flux in the stellar line cores. The RM effect was not corrected for the lines presented hereafter, as it is always encompassed inside the stellar line cores, and we verified that excluding these regions when building the transmission spectrum does not change the results. An example of these two effects is presented in Sect.\,\ref{Sub_sub_sec_Na}.

\subsubsection{The sodium doublet}\label{Sub_sub_sec_Na}
\begin{figure*}[h]
\resizebox{\hsize}{!}{\includegraphics[height=\textheight]{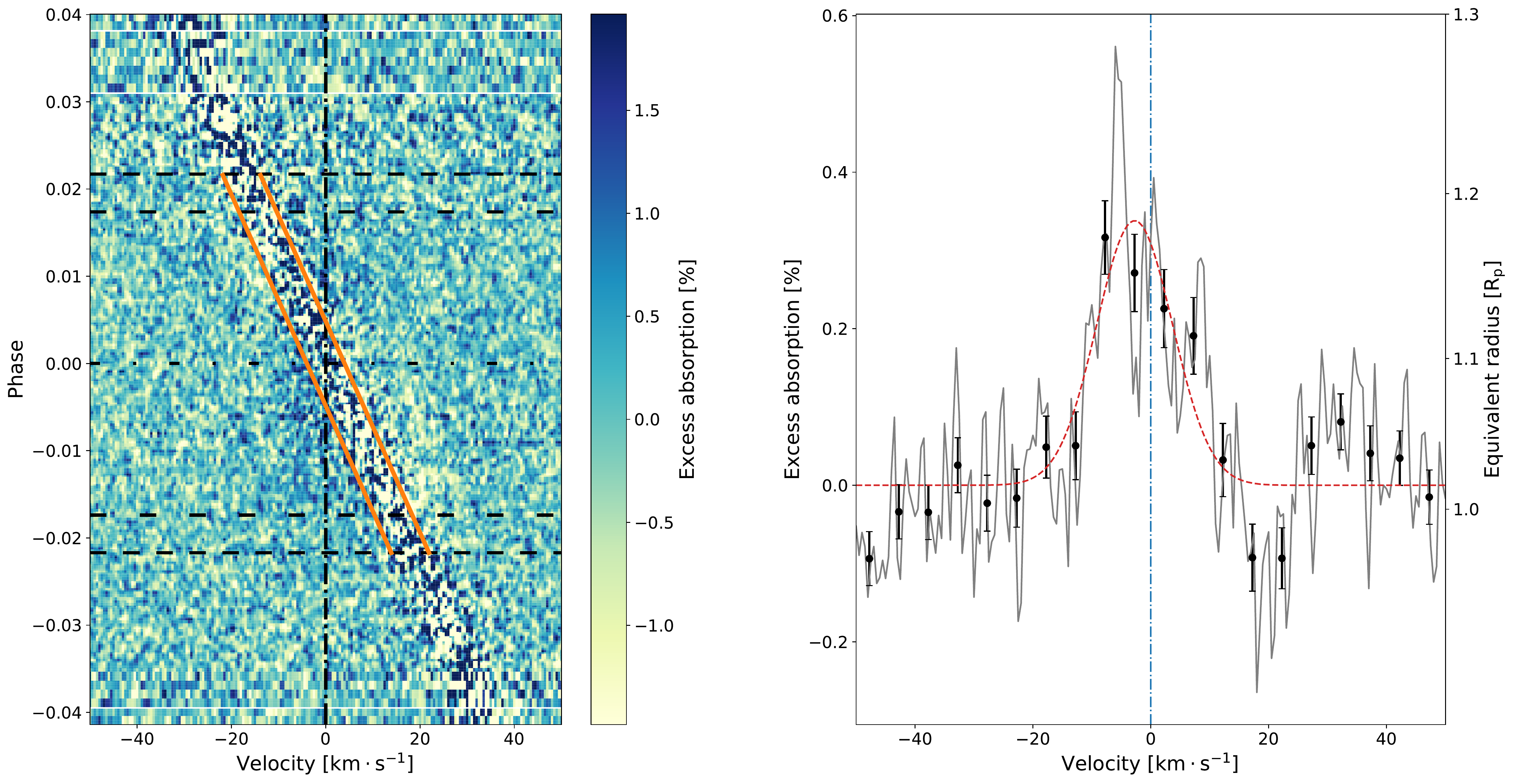}}
\caption[measured_CCF_Na_and_map]{\textit{Left:} Co-addition of the Na doublet lines as function of orbital phase in the planet rest frame considering the two transits. The vertical black dashed dotted lines represents the expected position of planetary sodium. The two horizontal black dashed lines are the t$_{1}$ and t$_{4}$ contact points while the loosely dashed lines are the t$_{2}$ and t$_{3}$ contact points. The two orange lines encompass the FWHM of the stellar local lines causing the RM effect. The sodium stellar line core low SNR is clearly visible at any phases. \textit{Right:} Co-addition of the \ion{Na}{i} doublet averaged across the two transits in grey and binned by ten elements in black. The vertical blue dash dotted line represents the expected position of planetary sodium. The red curve is the best gaussian fit on the data.}
\label{measured_CCF_Na_and_map}
\end{figure*}

\begin{figure}[h]
\resizebox{\hsize}{!}{\includegraphics[height=\textheight]{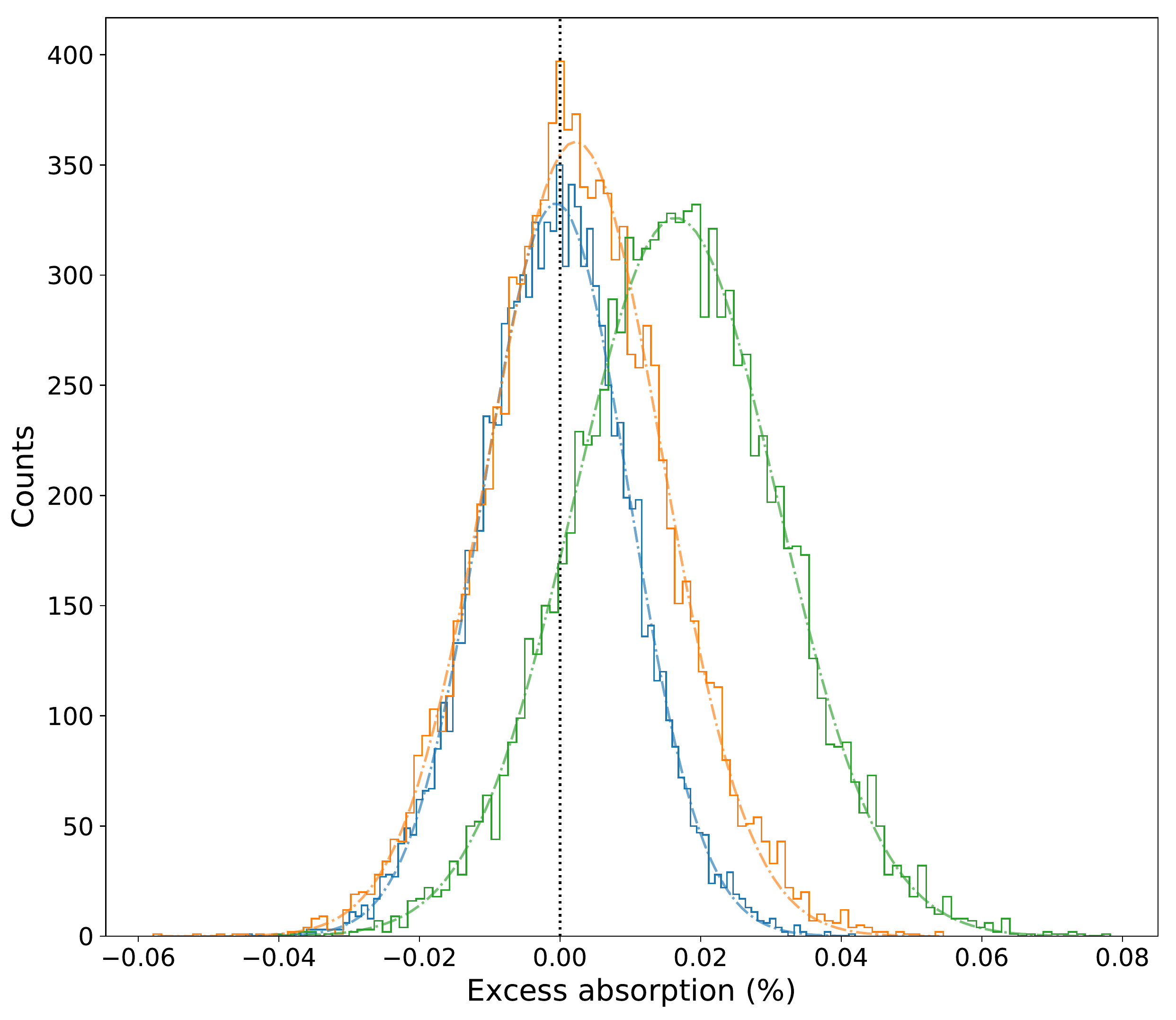}}
\caption[Na_EMC_star]{EMC simulations for the co-addition of the sodium doublet. Three distributions have been randomized 10000 times each: in blue the in/in distribution, in orange the out/out distribution and in green the in/out distribution. As expected, only the in/out distribution exhibits excess absorption.}
\label{Na_EMC_star}
\end{figure}

\begin{figure*}[h]
\resizebox{\hsize}{!}{\includegraphics[height=\textheight]{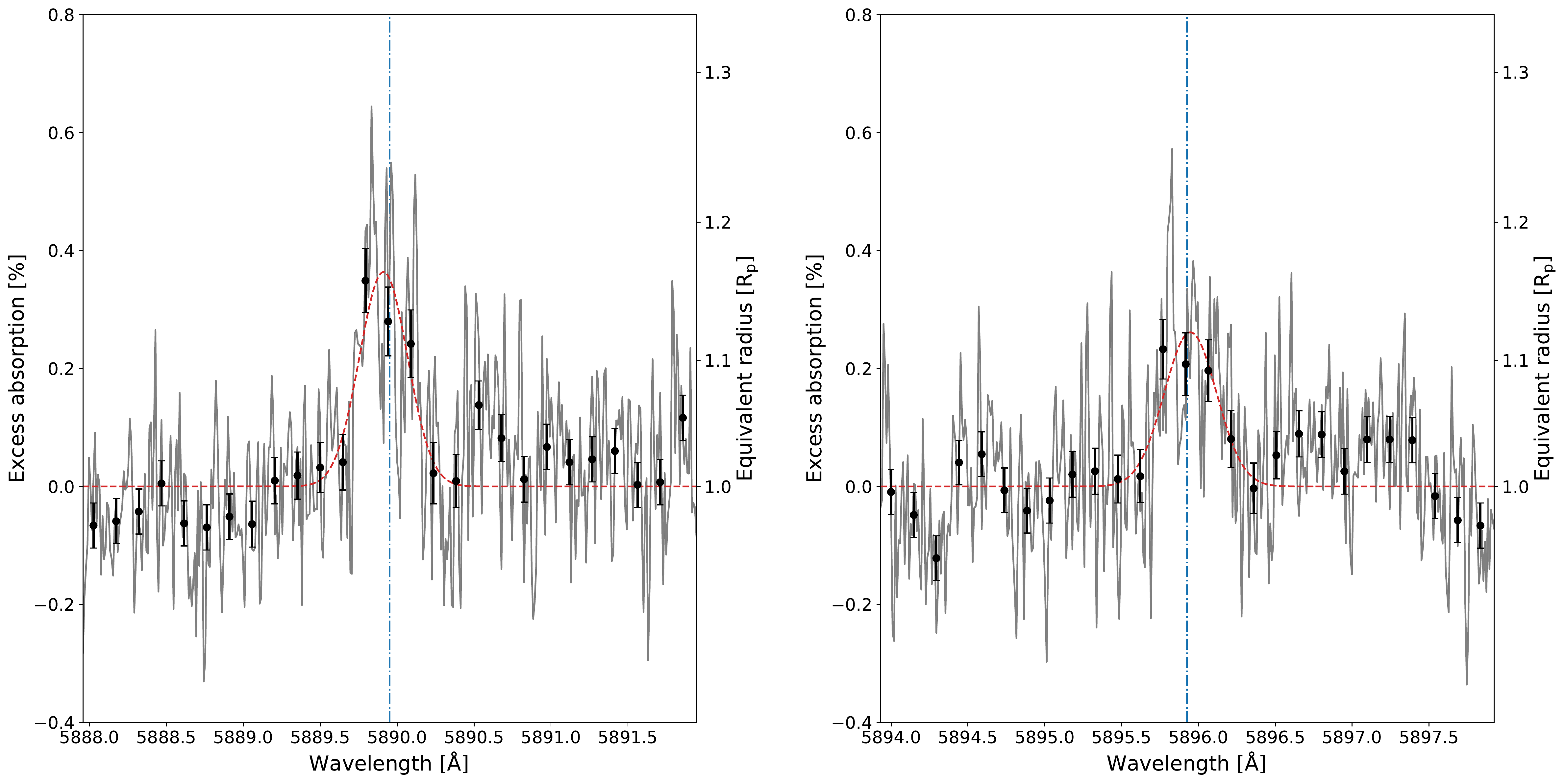}}
\caption[measured_TS_Na_doublet]{Transmission spectrum around the Na\,D2 (\textit{left}) and Na\,D1 (\textit{right}) line averaged across the two transits in grey and binned by fifteen elements in black. The vertical blue dash dotted line represents the expected position of planetary sodium lines.}
\label{measured_TS_Na_doublet}
\end{figure*}

Fig.\,\ref{measured_CCF_Na_and_map} (left panel) shows the evolution of the co-added sodium doublet lines in velocity space in the planet rest frame. We binned on this map the two transits. One can see the individual transmission spectra of each night at the lower and higher phases where the two datasets do not overlap. The low SNR of the stellar sodium line cores is also clearly visible as a noisy band showing both excess emission (blue) and excess absorption (orange) following the stellar track \citep{borsa_stellar_2018}. Such an effect cannot be mistaken with any other as it is a high-frequency noise that happens not only during transit but at all phases. The expected impact of the RM effect is encompassed by the white lines that follow the local stellar velocity in the planet's rest frame.  Moreover, due to the slow change of stellar velocity across the transit chord, the RM effect is only present inside the low S/N regions of the stellar sodium line cores, as discussed previously. In addition, one can also see some broad excess absorption following the sodium planetary track in blue. Fig.\,\ref{measured_CCF_Na_and_map} (right panel) shows the transmission spectrum of the co-added sodium lines in the planet rest frame. Excess absorption is visible and Gaussian fits yield an excess absorption of 0.34\,$\pm$\,0.04\,\% (9-$\sigma$) with a blueshift of 2.74\,$\pm$\,0.79\,km$\cdot$s$^{-1}$ and a FWHM of 15.18\,$\pm$\,1.75\,km$\cdot$s$^{-1}$. This excess absorption corresponds to an extension of the atmosphere of 1.15\,R$_p$ (6.8 scale heights), which is well below the Roche Lobe radius (1.96\,R$_p$).\\ 
To validate the significance of this signal, we performed Empirical Monte Carlo (EMC) simulations. We followed the method described in \cite{wyttenbach_spectrally_2015,casasayas-barris_atmospheric_2019,seidel_hot_2019} that consists of generating three scenarios (in/in, out/out and in/out) of 10 000 iterations each. For one iteration, the in- and out-of-transit spectra are randomized among the pool of spectra for the considered scenario. The absorption was measured on a 2\,\AA\ passband centered on both lines on the transmission spectrum. The results are shown in Fig.\,\ref{Na_EMC_star}. As expected, the in/in and out/out scenarios are centered at zero absorption while the in/out scenario shows excess absorption. These EMC simulations confirm that the absorption signal is not serendipitous but linked to the planet atmosphere.\\
As the co-addition of the two sodium lines has a detected planetary signal, we have a look at the individual lines in Fig.\,\ref{measured_TS_Na_doublet} for  \ion{Na}{i}\,D1 and  \ion{Na}{i}\,D2. Both lines are compatible with respectively excess absorption of 0.26\,$\pm$\,0.05\,\%  and 0.36\,$\pm$\,0.06\,\% , radial velocity shift of 1.19\,$\pm$\,1.60\,km$\cdot$s$^{-1}$ and -2.02\,$\pm$\,1.31\,km$\cdot$s$^{-1}$, and a FWHM of 20.43\,$\pm$\,3.46\,km$\cdot$s$^{-1}$ and 18.70\,$\pm$\,2.80\,km$\cdot$s$^{-1}$.
 
\subsubsection{Other atmospheric tracers: K, Li and H-$\alpha$}\label{Sub_sub_sec_other_species}
\begin{figure}[h]
\resizebox{\hsize}{!}{\includegraphics[height=\textheight]{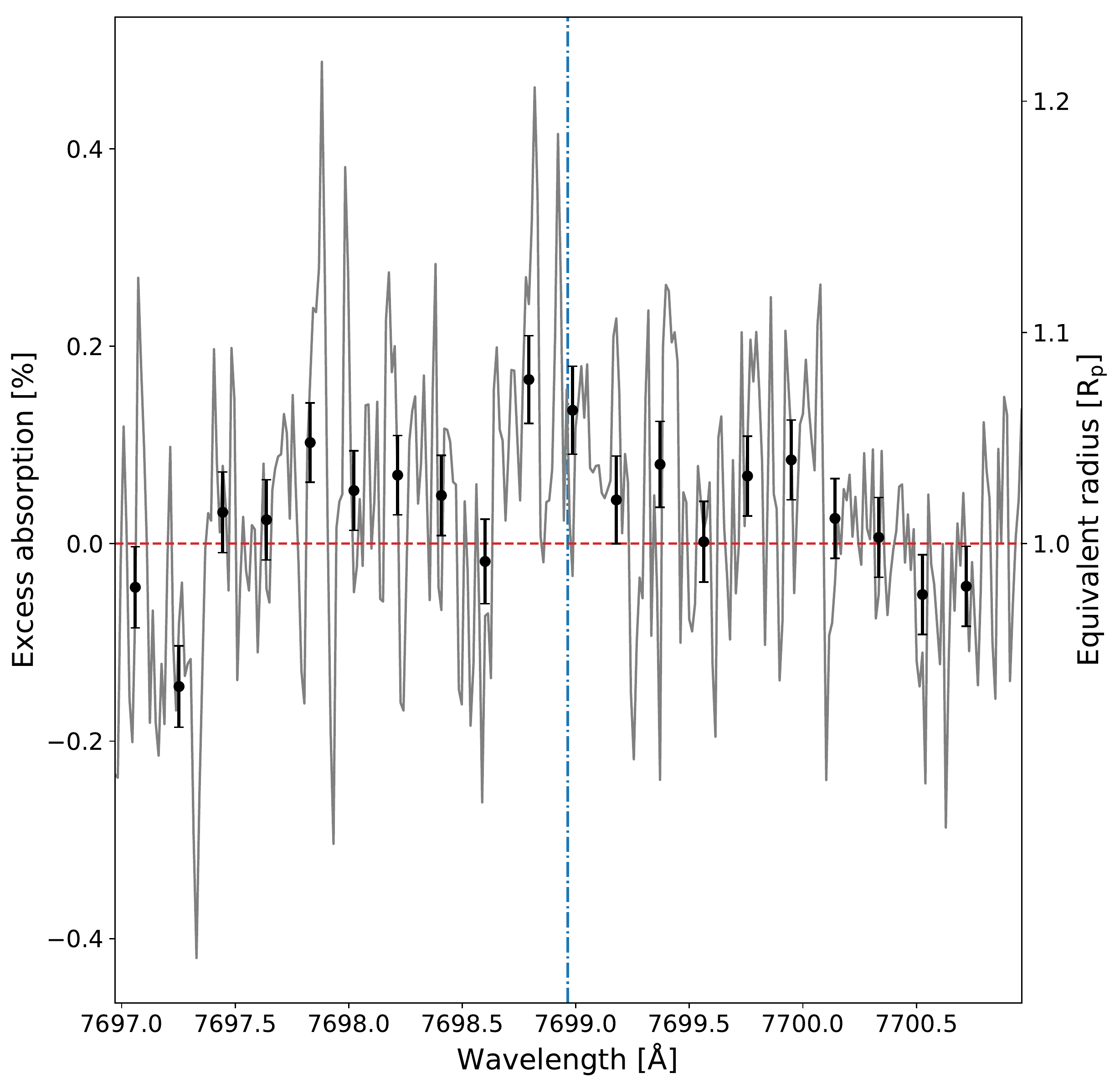}}
\caption[measured_TS_K]{Same as figure\,\ref{measured_TS_Na_doublet} for the potassium line (7698.96\,\AA).}
\label{measured_TS_K}
\end{figure}

\begin{figure}[h]
\resizebox{\hsize}{!}{\includegraphics[height=\textheight]{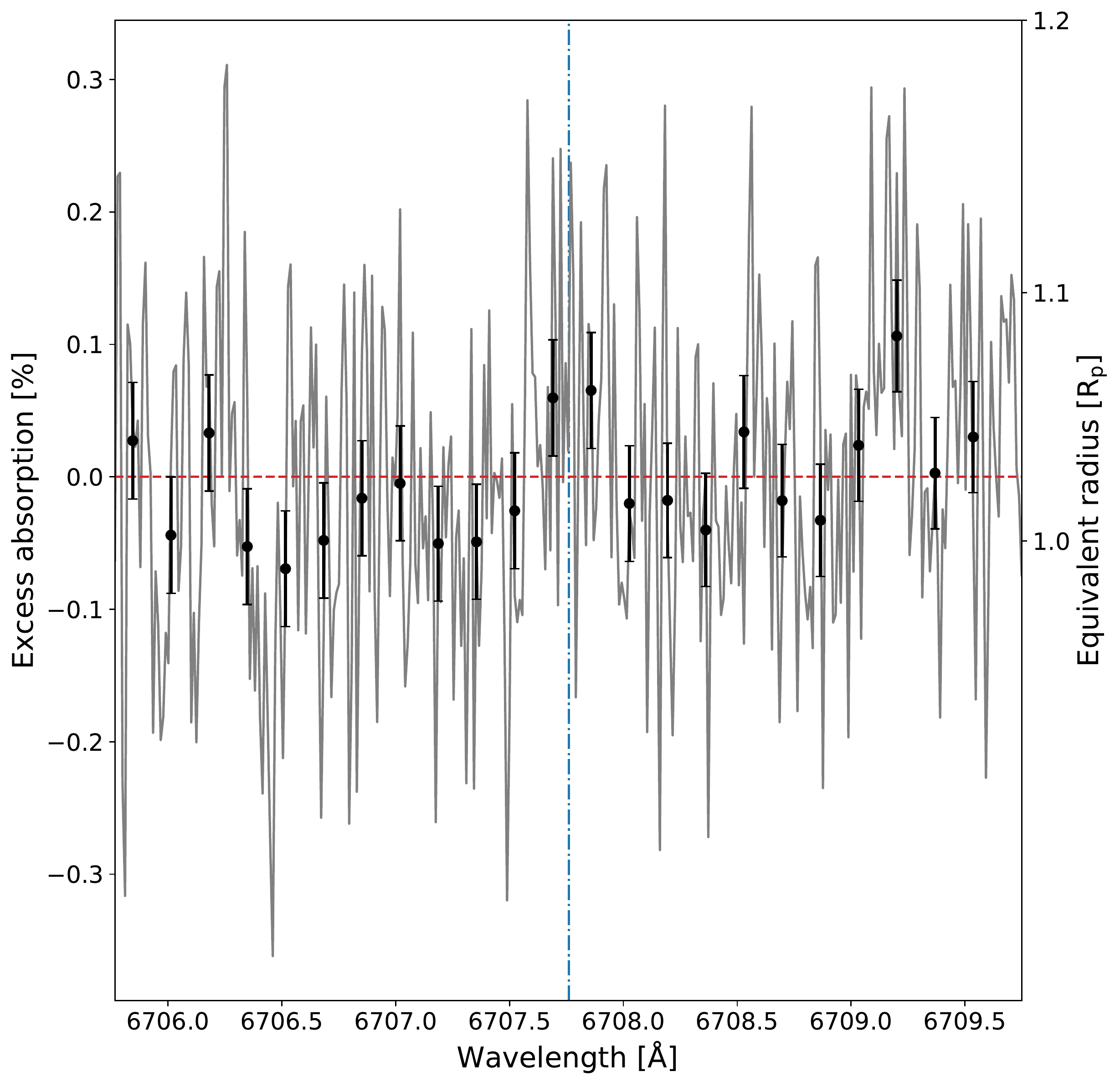}}
\caption[measured_TS_Li]{Same as figure\,\ref{measured_TS_Na_doublet} for the lithium line (6707.76\,\AA).}
\label{measured_TS_Li}
\end{figure}

\begin{figure}[h]
\resizebox{\hsize}{!}{\includegraphics[height=\textheight]{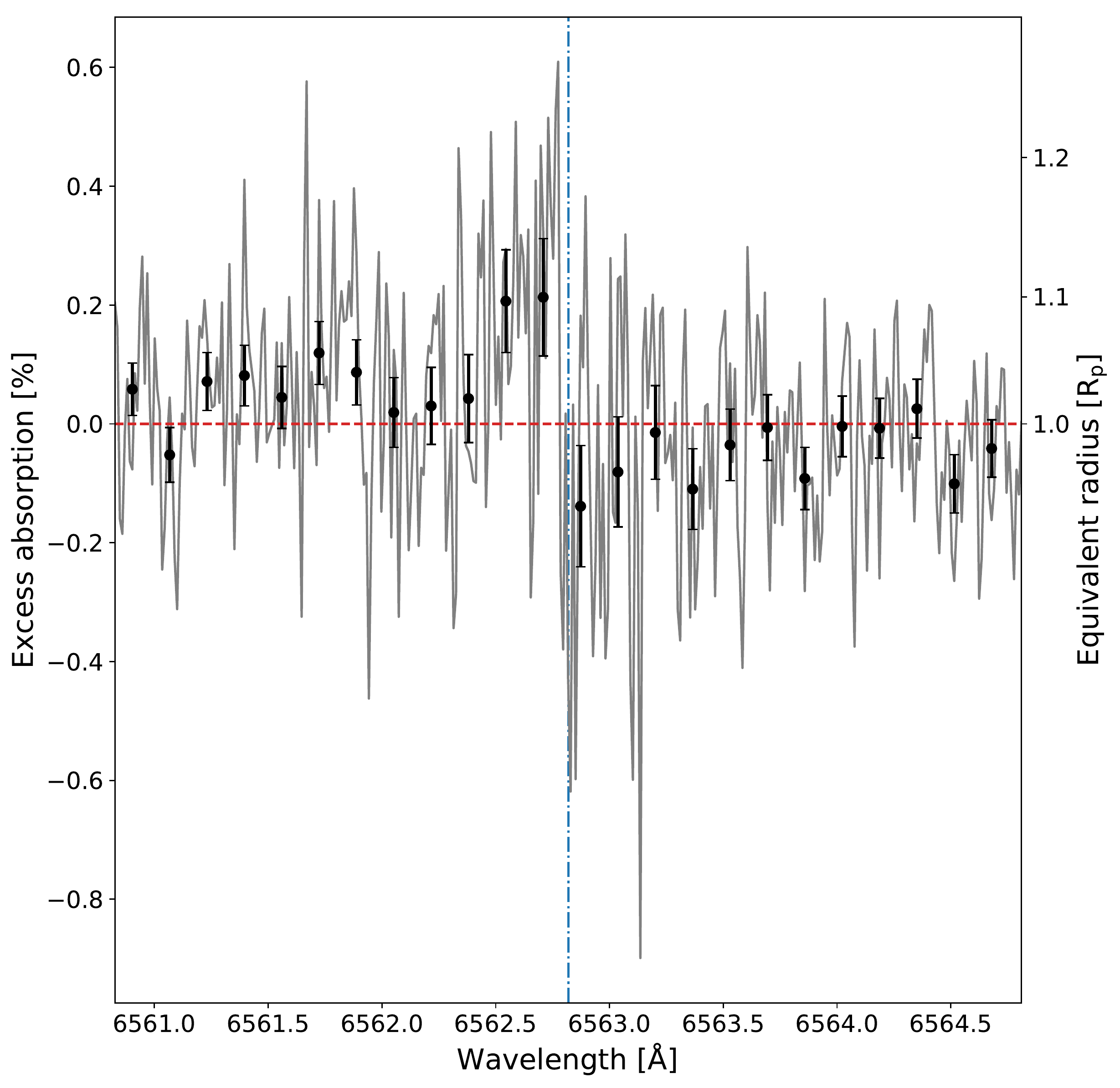}}
\caption[measured_TS_Ha]{Same as figure\,\ref{measured_TS_Na_doublet} for the H-$\alpha$ line (6562.82\,\AA).}
\label{measured_TS_Ha}
\end{figure}

We also searched for the potassium D1 line at 7698.96\,\AA\ (the \ion{K}{i}\,D2 line was not studied due to strong O$_2$ telluric contamination), the lithium line at 6707.76\,\AA\ and the H-$\alpha$ line at 6562.82\,\AA. Similarly to the \ion{Na}{i} doublet, these lines are affected by low SNR residuals in the stellar line core. In Figs.\, \ref{measured_TS_K}, \ref{measured_TS_Li} and \ref{measured_TS_Ha}, we present the combined transmission spectrum around each line. However, we do not detect any of the species even if the 1-pixel dispersion is respectively of 0.14, 0.12, and 0.20\,\%. Assuming a line width of 10\,km$\cdot$s$^{-1}$ and following the formalism of \cite{allart_search_2017}, we can exclude excess absorption from these atmospheric tracers at a confidence level of 0.09, 0.08, and 0.13\,\% at 3-$\sigma$. These upper limits correspond to 1.04, 1.04, and 1.06\,R$_p$ or 1.9, 1.7, and 2.8 scale height (H).

\subsection{Search for water vapor}\label{Sub_sec_Water}
\begin{figure}[h]
\resizebox{\hsize}{!}{\includegraphics[height=\textheight]{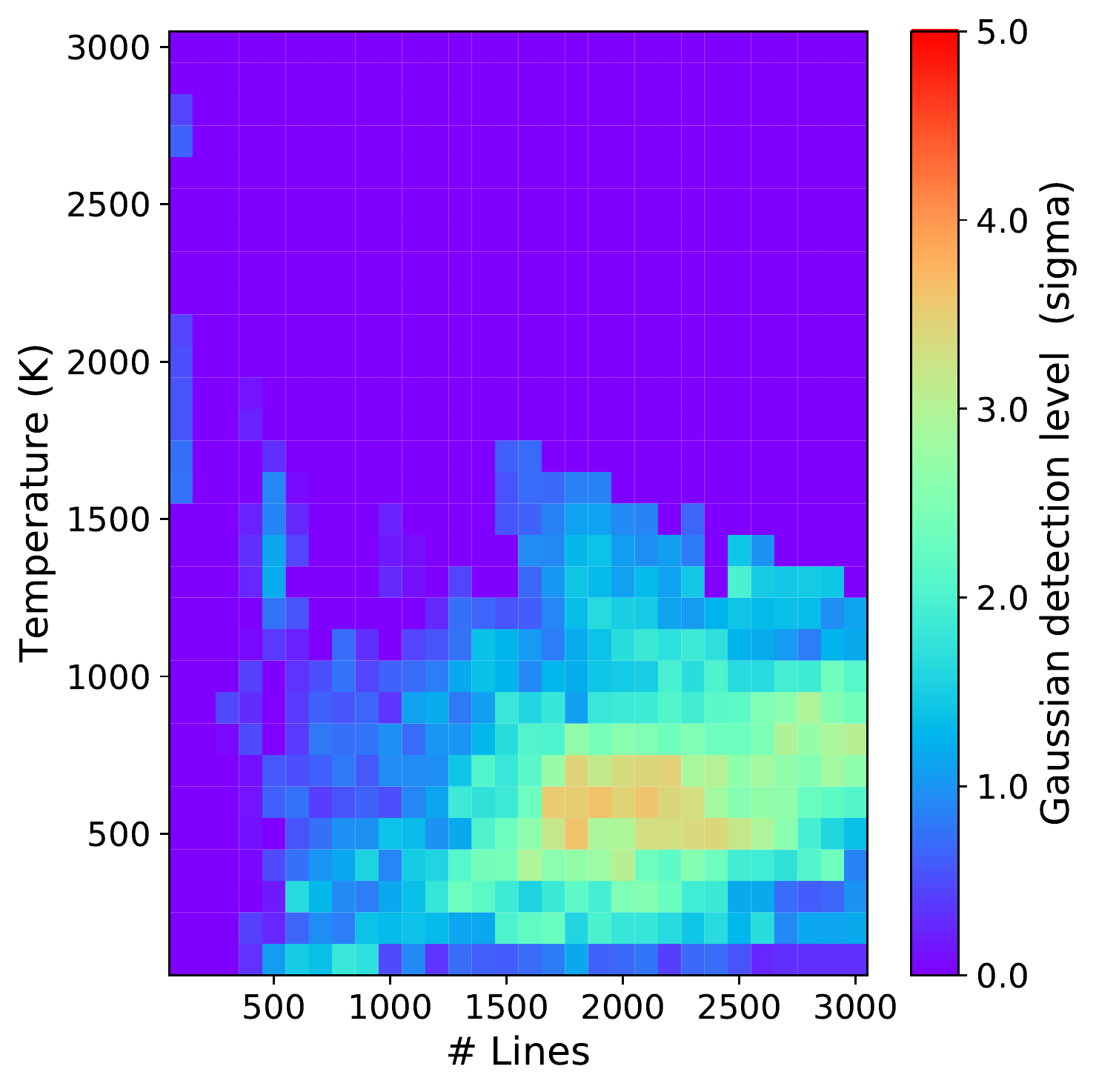}}
\caption[Injection_test_best_fit]{Detectability map based on a gaussian fit for a set of binary CCFs with temperature ranging from 100 to 3000\,K and including from the 100 to 3000\,strongest water lines.}
\label{HITEMP_detectability}
\end{figure}

We applied the method described in Sect\,\ref{Sub_sec_CCF}, and we show in Fig.\,\ref{HITEMP_detectability} the detection levels obtained from a Gaussian fit applied to the set of CCFs with a broad range of temperature and number of lines. All CCFs have gaussian features with a confidence level at least lower than 3.6-$\sigma$. We can thus conclude that there is no robust detection of water vapor in the visible.

\begin{figure*}[h]
\includegraphics[width=\textwidth]{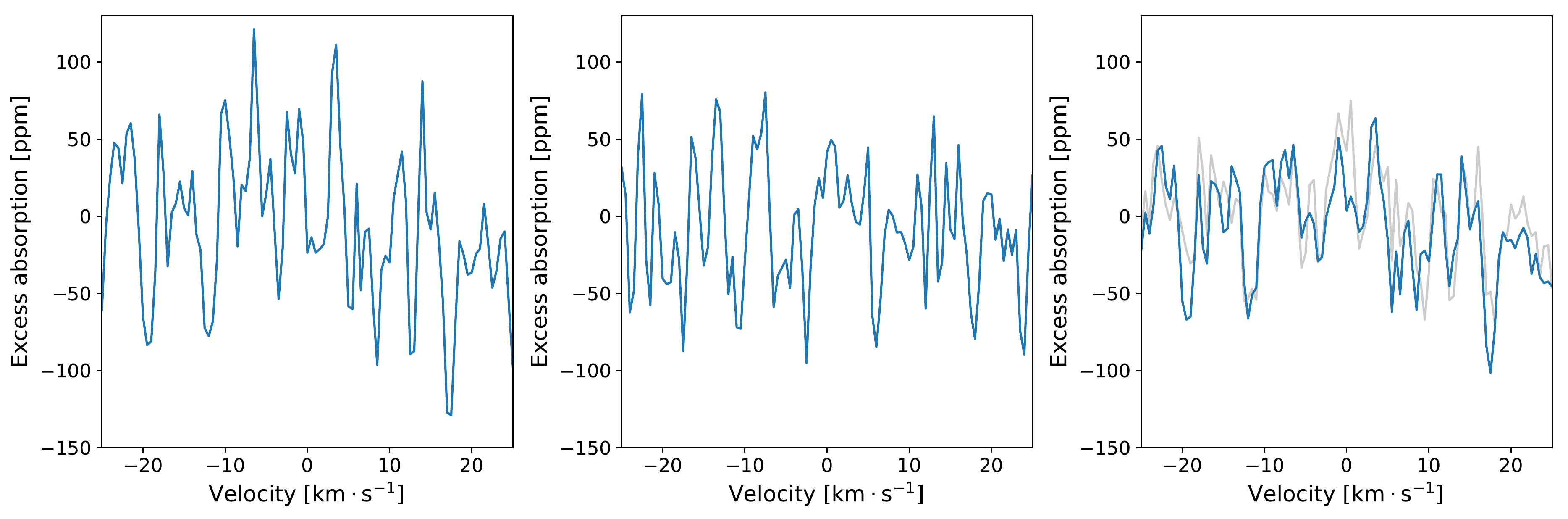}
\caption[Water_1400K_1600lines]{Measured CCFs in the planet rest frame with water vapor mask at 1400\,K and 1600\,lines for both nights (\textit{left panel}: 24-02-2020 and \textit{middle panel}:17-03-2020) and the combination of them (\textit{right panel}). On the \textit{right panel}, the combined CCF with the highest confidence level from the temperature vs. number lines map (600\,K and 1900\,lines; Fig.\,\ref{HITEMP_detectability}) has been added in grey.}
\label{Water_1400K_1600lines}
\end{figure*}

To constrain the presence of water vapor and based on the equilibrium temperature and the results of Appendix\,\ref{AppendixC}, we looked at the CCF built with the mask with 1400\,K and 1600\,lines. This is shown in Fig.\,\ref{Water_1400K_1600lines} for both nights and the average. We added for comparison purposes the CCF, from Fig.\,\ref{HITEMP_detectability}, showing the highest confidence level in grey obtained with the mask at 600\,K and 1900\,lines. The lack of difference between both CCFs reinforce the claim of non-detection.\\
For the CCF at 1400\,K and 1600\,lines, the 1-pixel dispersion in the continuum for both nights is respectively of 41 and 57\,ppm and for the combined CCF of 29\,ppm. Based on Appendix\,\ref{AppendixC}, we can assume a line width of 2.5\,km$\cdot$s$^{-1}$ leading to a 1-$\sigma$ precision of 13\,ppm and thus to a 3-$\sigma$ upper limit of 38\,ppm which corresponds to 0.08\,H. We described the impact of this exquisite precision in Sect.\,\ref{Sec_interp}.

\section{Combining low- and high-resolution spectroscopy}\label{Sec_interp}
Table\,\ref{Summary_detections} summarizes the detection and non-detection obtained with the ESPRESSO data. In this section, we will first reconciliate the low-resolution \textit{HST} results to our high-resolution results and then discuss in more details the sodium detection.
\begin{table}[h]
\centering
\caption{Atmospheric detection and non-detections summary for WASP-127b with ESPRESSO}
\begin{tabular}{lccc}
\hline
\textit{Detection}  & Absorption & Equivalent R$_p$ & H \\
\hline
Na doublet & 0.34\,$\pm$\,0.04\,\% &1.15 & 6.8 \\
\hline
\\
\hline
\textit{Non-detection } & 3-$\sigma$ upper limit & Equivalent R$_p$ & H \\
\hline
K D1 & 0.09\,\% &1.04 &1.9 \\
Li & 0.08\,\% &1.064 &1.7 \\
H-$\alpha$ & 0.08\,\% &1.04 &2.8 \\
H$_2$O & 38\,ppm &1.002 &0.08 \\
\hline
\end{tabular}
\label{Summary_detections}
\end{table}

ESPRESSO allows us to reach the fine 1-$\sigma$ precision of 13\,ppm on the search of water vapor in the visible, which translates into 3-$\sigma$ upper limit 0.08\,H. This unique precision can be used to largely constrain the presence of clouds in the exoplanet's atmosphere by studying several water bands \citep{pino_diagnosing_2018,alonso-floriano_multiple_2019,sanchez-lopez_water_2019}. Therefore, to see if the water detection at 1.3 microns obtained at low-resolution is compatible with the non-detection obtained in the visible at high-resolution, we computed a set of high-resolution models described in Sect\,\ref{model_description} to match both results.

\subsection{Model description}\label{model_description}
To interpret our results we employed the $^\pi \eta$ line-by-line radiative transfer code \citep{ehrenreich_transmission_2006,ehrenreich_transmission_2012,pino_combining_2018}. This code was already applied for the simultaneous interpretation of ground-based, optical, high spectral resolution observations with HARPS and \textit{HST}/WFC3 observations \cite{pino_diagnosing_2018}. Here, we employ the version by \citep{pino_diagnosing_2018}, who extended $^\pi \eta$ to simulate the water bands observed by multiple optical to near-IR high spectral resolution instruments including ESPRESSO, and is thus suitable to our goal.\\
The considered opacity sources include sodium, potassium, water, Rayleigh scattering by molecular hydrogen, collisional-induced absorption (CIA) and aerosols distinguished between a gray absorber and a chromatic absorber. We name those "clouds'' and "hazes'' respectively, following the convention by Earth atmosphere scientists. The alkali doublets are modelled with a Voigt profile following \cite{burrows_near-infrared_2000, iro_time-dependent_2005}. For water, we employ the HITEMP line list \cite{rothman_hitran2012_2013}, converted to a cross-section with the HELIOS-K code \citep{grimm_helios-k_2015}. The Rayleigh scattering cross-section for molecular hydrogen follows \cite{dalgarno_rayleigh_1962,dalgarno_properties_1965}, and the CIA cross-section follows \cite{richard_new_2012}. Gray aerosols are modelled by setting the atmosphere as opaque below a pressure $P_\mathrm{c}$. Finally, the cross-section of chromatic absorption by aerosols ($\mathrm{\sigma_h}$) is modelled as a power law of the form:
\begin{equation}
\sigma_h = S\cdot\left(\frac{\lambda}{3,500}\right)^{-4}\,,
\end{equation}
where $\lambda$ is the wavelength in Angstroms and $S$ is a scale factor normalized to the Rayleigh scattering cross-section of molecular hydrogen at $3500~\mathrm{\AA}$. Gray absorbers dominate the aerosol opacity budget at the wavelengths of our interest covered by ESPRESSO and \textit{HST} WFC3, thus we consider our simple model adequate for our purpose. \cite{pino_combining_2018} provide a detailed description of the opacities and the assumptions in the model. \\
A full joint low-resolution (LR) + high-resolution (HR) retrieval is out of the scope of this paper. We instead consider an isothermal atmosphere, fix the abundances of sodium, potassium and water to the free-free retrieval results by \cite{spake_abundance_2020}, and produce a grid of models with different cloud coverage. Indeed, for a given abundance and reference pressure/radius, it is aerosol coverage that controls the relative strength of water features \citep{de_kok_identifying_2014,pino_diagnosing_2018}, which is our observable at high spectral resolution. We summarize in Tab.\,\ref{table_model} the parameters employed in our grid of models.

\begin{table*}
\caption{Input $^{\pi} \eta$ parameters used in our model grid.}\label{table_model}
\begin{tabular}{cccc}
\hline 
Parameter & Value & Definition & Notes \\ 
\hline 
$R_s [R_\odot]$ & 1.39 & Stellar radius & \cite{lam_dense_2017} \\ 

T [K] & 820 & Atmospheric temperature & \cite{spake_abundance_2020} Tab.\,11 \\ 

$R_p(10~\mathrm{bar}) [R_j]$ & 1.3586 & Planet radius at 10 bar level & Corresponds to 1.4 at 1\,bar, consistent with \cite{spake_abundance_2020}, Tab.\,11 \\ 

$M_p [M_j]$& 0.18 & Planet mass &  \cite{lam_dense_2017} \\ 

$S$ & 2.59 & Hazes scale factor & does not significantly impact the strength of the water features\\ 

 &  &  & probed by ESPRESSO and \textit{HST}, see text \\ 

$\log_{10} \mathrm{VMR_{Na}}$ [cgs] & -4.89 & Volume mixing ratio Na & \cite{spake_abundance_2020} Tab.\,11 \\ 

$\mathrm{VMR_{K}}$ & -8.01 & Volume mixing ratio K & \cite{spake_abundance_2020} Tab.\,11 \\ 

$\mathrm{VMR_{H_2O}}$ & -2.87 & Volume mixing ratio water & \cite{spake_abundance_2020} Tab.\,11 \\ 

$P_\mathrm{c}$ [bar] & $10^{4}$--$10^{-5}$ & Gray clouds top & Sampled in logarithmically spaced steps \\ 
\hline 
\end{tabular} 
\end{table*}

\subsection{Water vapor and cloud deck}
\begin{figure*}[h]
\includegraphics[width=\textwidth]{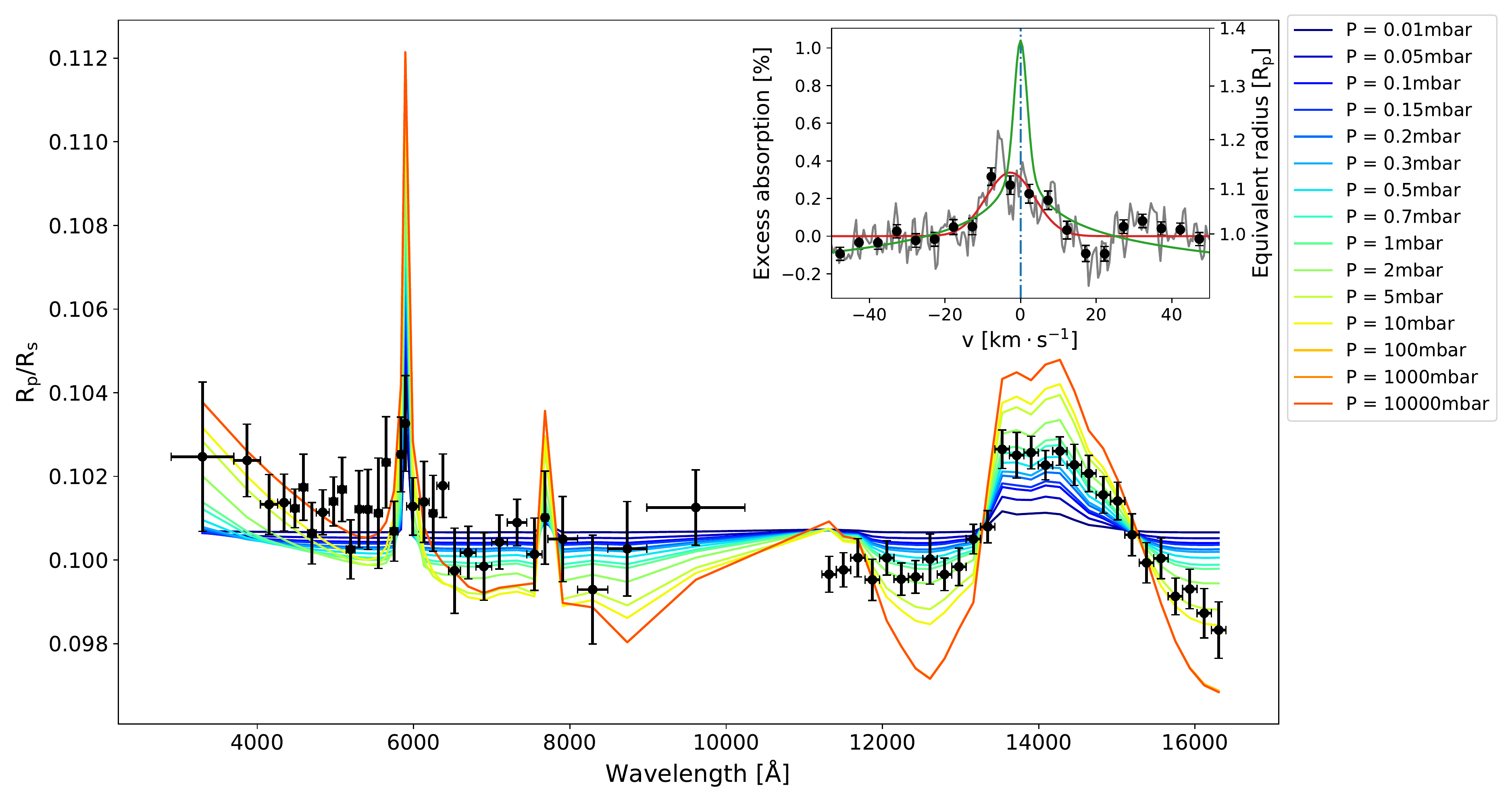}
\caption[LR_models]{The low-resolution model described in Sect.\,\ref{model_description} with different cut in pressure to represent various gray cloud decks (colored curves). In black, the \textit{HST} data from \cite{spake_abundance_2020}. The inset is the same as Fig.\,\ref{measured_CCF_Na_and_map} (right panel) where we added in green the co-addition of the high-resolution model Na doublet lines which is not affected by the cloud deck pressure.}
\label{LR_models}
\end{figure*}

\begin{landscape}
\begin{figure}
\centering
\includegraphics[width=1.3\textwidth]{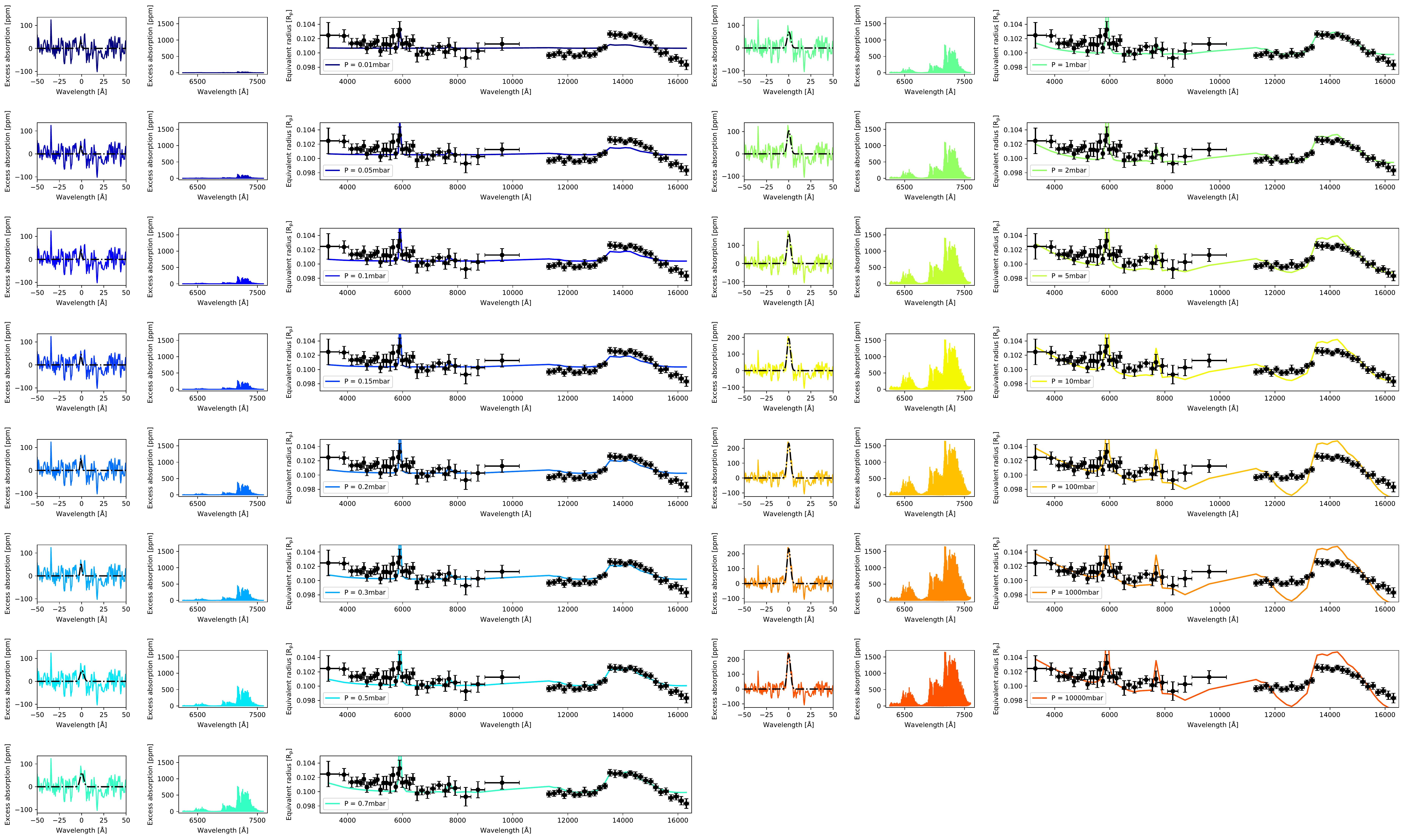}
\caption[Summary_fit_eta]{Combining low and high-resolution datasets. The color-code is the same as figure\,\ref{LR_models}. \textit{Left column}: ESPRESSO CCFs with injected models retrieved for the equilibrium temperature and 1600 lines. The best gaussian-fit model shown in black evaluates if the injected model could be retrieved in the ESPRESSO data. \textit{Middle column}: The high-resolution model that is injected into the ESPRESSO data. \textit{Right column}: The \textit{HST} data in black are compared to low-resolution models. A minimization algorithm is applied to estimate the $\chi^2_{red}$ and BIC of each model compared to the data. \textit{Rows} correspond to the same model with different cut in pressure simulating different gray cloud-deck.}
\label{Summary_fit_eta}
\end{figure}
\end{landscape}

We first convolved the models to the \textit{HST} resolution and binned the models to match the \cite{spake_abundance_2020} data sampling. We then adjusted each LR model to these data by fitting a constant vertical offset. The $\chi^2_{red}$ and BIC values have been reported in Table\,\ref{Fit_low_high}. Figs\,\ref{LR_models} and \ref{Summary_fit_eta} shows those models and the \textit{HST} data.\\
Concomitantly, we convolved the high-resolution model (hereafter HR models) by the instrumental LSF (2.0\,km$\cdot$s$^{-1}$) of ESPRESSO and the planetary rotation (1.7\,km$\cdot$s$^{-1}$ assuming tidally-lock rotation). We resampled the HR models to match the ESPRESSO sampling. Then, we only consider the spectral range between 6250 and  7600\,\AA\, for which we removed the continuum to render the HR models as they should be observed from the ground, where the continuum is lost. Those excess absorption HR models are shown in the middle column of Fig.\,\ref{Summary_fit_eta}. Hereafter, we injected those HR models into the ESPRESSO data, and we computed the CCFs with a binary mask at T$_{eq}$ and 1600\,lines. Finally, we fitted a Gaussian to each of those CCFs to measure if the injected signal can be retrieved. We report in Table\,\ref{Fit_low_high}, the Gaussian amplitude and the detection level.\\
Based on this analysis summarized in Fig.\,\ref{Summary_fit_eta} and Table\,\ref{Fit_low_high}, we cannot only see if one model can reproduce both LR infrared and HR visible result but also what constraint the ESPRESSO data can bring to \textit{HST}/WFC3 data. On the LR side, we see that the cloud-free (P$_c$>100\,mbar) and the most cloudy models do not fit well the \textit{HST}/WFC3 data, but there are, in between, a range of models that do. The best model is obtained for pressure of 0.5\,mbar, but models with pressure ranging from 0.3 to 1\,mbar are equally good as their $\Delta$BIC is lower than 6. On the HR side, we can see that a clear signal is detected for the cloud-free models (as expected in Appendix \ref{AppendixC}), and thus those models do not reproduce the ESPRESSO results (Sect\,\ref{Sub_sec_Water}). Models for which the injected signal is recovered at more than 3-$\sigma$ (i.e., a hint of detection) can be rejected. Therefore, our ESPRESSO data are compatible with grey clouds that have a pressure below 0.5\,mbar. Combining the LR and HR data allows us to constrain the cloud deck pressure between 0.3 and 0.5\,mbar. Based on the best model (P=0.3\,mbar), we estimate that the 1.3 microns water band at high-resolution extend up to 6 scale height. This is the first time that high-resolution visible data is used to constrain the presence of clouds using water vapor lines. It is important to remind that there is a degeneracy with water abundance and reference pressure, and that we chose only the best fit values from \cite{spake_abundance_2020}.

\begin{table*}[h]
\centering
\caption{Grey cloud exploration through injection tests to combine low- and high-resolution datasets. The model highlighted in blue is the best model once LR and HR data are combined. Models in orange are the models compatible with LR data only, while models in olive are compatible at the 90\,\% confidence-level with HR data only.}
\begin{tabular}{*{5}{>{\rowfonttype}c}<{\rowfont{}}}
\hline
Pressure [mbar]  & \textit{HST} $\chi^2_{red}$ & \textit{HST} BIC & ESPRESSO retrieved amplitude & ESPRESSO retrieved significance\\
\hline
\rowfont{\color{olive}} 0.01  &  3.47  &  212.3  &  35 \,$\pm$\, 23  &  1.5 \\ 
\rowfont{\color{olive}} 0.05  &  2.76  &  169.45  &  ---  &  0.0 \\ 
\rowfont{\color{olive}} 0.1  &  2.39  &  147.44  &  43 \,$\pm$\, 22  &  1.9 \\ 
\rowfont{\color{olive}} 0.15  &  2.28  &  140.83  &  45 \,$\pm$\, 22  &  2.0 \\ 
\rowfont{\color{olive}}  0.2  &  2.09  &  129.67  &  36 \,$\pm$\, 14  &  2.5 \\ 
\rowfont{\color{blue}} 0.3  &  2.02  &  125.43  &  39 \,$\pm$\, 14  &  2.8 \\ 
\rowfont{\color{orange}} 0.5  &  1.94  &  120.44  &  49 \,$\pm$\, 14  &  3.5 \\ 
\rowfont{\color{orange}} 0.7  &  1.95  &  121.13  &  63 \,$\pm$\, 14  &  4.4 \\ 
\rowfont{\color{orange}} 1  &  2.0  &  124.02  &  71 \,$\pm$\, 14  &  5.0 \\ 
2  &  2.34  &  144.57  &  104 \,$\pm$\, 14  &  7.2 \\ 
5  &  3.41  &  208.68  &  162 \,$\pm$\, 15  &  11.2 \\ 
10  &  4.23  &  258.09  &  193 \,$\pm$\, 15  &  13.2 \\ 
100  &  7.94  &  480.43  &  233 \,$\pm$\, 15  &  16.0 \\ 
1000  &  7.97  &  482.39  &  233 \,$\pm$\, 15  &  16.0 \\ 
10000  &  7.97  &  482.39  &  233 \,$\pm$\, 15  &  16.0 \\ 
\hline
\end{tabular}
\label{Fit_low_high}
\end{table*}
 
\subsection{Atomic species}
High-resolution data from the ground cannot probe the wings of the line far from the core because of the loss of absolute flux level. However, for low-resolution spectrographs, the lines' cores are diluted over tens of angstroms. The lack of line core signal can be compatible with the presence of wings if process such as ionization are considered. The ESPRESSO data do not indicate the presence of potassium and lithium cores with an upper limit of 1.9 and 1.7\,H.  Potassium and lithium have only been reported on one LR dataset \citep{chen_gtc_2018} but not with \textit{HST} \citep{spake_abundance_2020}. Therefore, the situation is currently unclear with respect to potassium and lithium in the atmosphere of WASP-127b.\\

As presented in Sect.\,\ref{Sub_sub_sec_Na}, we report the detection of sodium in the atmosphere of WASP-127b at the 9-$\sigma$ level. We compare its absorption with LR datasets by convolving our transmission spectrum around the sodium to the resolution of \textit{HST}/STIS and by binning over 35\,\AA\ to fit the \textit{HST}/STIS bin centered at 5895\,\AA. Our ESPRESSO Na excess absorption of $\sim$0.34\,\% becomes over the 35\,\AA\ bin an excess absorption of $\sim$0.03\,\% which is consistent with the relative absorption between the \textit{HST} bin centered at 5895\,\AA\ and its two adjacent bins. Therefore, our high-resolution detection is compatible with the detection of \cite{spake_abundance_2020}, and also with \cite{chen_gtc_2018} who detected similar signal as \textit{HST}.\\
We also report a slight blueshift on the sodium core of 2.74\,$\pm$\,0.79\,km$\cdot$s$^{-1}$. Blueshifts are not unusual for lukewarm and hot gas giant exoplanets as it has been reported on several exoplanets \citep{snellen_orbital_2010,brogi_rotation_2016,allart_spectrally_2018, hoeijmakers_atomic_2018,flowers_high-resolution_2019, bourrier_hot_2020,ehrenreich_nightside_2020} and could be due to zonal winds from the day-to-night side. In the inset of Fig.\,\ref{LR_models}, we overplotted to our Na detection the high-resolution model normalized to the local continuum. Since we are looking at the very narrow core of the planetary lines, the pressure cut applied does not impact the doublet line depth and all our models are thus similar. While our model reproduces the \textit{HST}/STIS data as we assumed the same Na abundance as \cite{spake_abundance_2020}, the shape is different at high-resolution. Our model has an expected Na signature of about 1\,\% which do not reproduce well the observed excess absorption. This difference might be due to a lack of hazes or an inappropriate Na abundance in our model. Moreover, the FWHM is not compatible either and is degenerated with many parameters such as Na abundance, hazes, various wind patterns, or thermal broadening (e.g., \cite{seidel_wind_2020}). In order to explain how such a small extension is possible, a dedicated modeling of this signature is needed.

\section{Conclusion}\label{Sec_concl}
We analyzed two transits of WASP-127b observed with ESPRESSO. We detected for the first time its RM effect and concluded that the planet has a misaligned retrograde orbit. We searched for the presence of atomic species, and despite exquisite precision of a few scale heights, we did not detect K, Li, or H-$\alpha$. Nonetheless, we confirmed the presence of sodium by measuring an excess absorption arising from the Na doublet core lines, which extends up to 7\,H only. This Na detection is compatible with the ones reported in the literature at low resolution \citep{chen_gtc_2018,spake_abundance_2020}.  In comparison, the water band at 1.3\,microns extend up to 6\,H if it was seen at high-resolution. Therefore, Na and H$_2$O are probing the same atmospheric layers which is quite uncommon for the gas giant planets. \\
We proposed a new framework to search for water vapor and other molecular species at high-resolution. This framework is based on Occam's razor principle to minimize the model-dependent variables. We tested it with success by injecting a model into the data, and then we applied it to the ESPRESSO data. Unfortunately, despite a good precision on the data, we did not detect water vapor, but we fixed an upper limit of 38\,ppm on the presence of water in the 700-750\,nm passband. Finally, we combined this result to low-resolution detection of water at 1.3\, microns to constrain the presence of clouds in the atmosphere of WASP-127b.\\

To conclude, we report for the first time that high-resolution visible data can differentiate between cloudy and cloud-free worlds by measuring the water content and can also provide essential information on the cloud deck pressure. The framework to measure this water content will be applied to other exoplanets in the ESPRESSO GTO atmospheric survey and on other surveys such as the NIRPS GTO atmospheric survey. One of the primary targets for this last survey could be, in fact, WASP-127b, where the near-infrared water excess absorption would be about 800-1000\,ppm once the telluric saturation bands are excluded.


\begin{acknowledgements}
We thank the referee for their careful reading and their inputs. The authors acknowledge the ESPRESSO project team for its effort and dedication in building the ESPRESSO instrument. We thank Jessica Spake for the discussion on the HST datasets and important insight.  We thank Monika Lendl and Julia Seidel for their important feedback and discussions on the EulerCam and HARPS data. We acknowledge the Geneva exoplanet atmosphere group for fruitful discussions. This work has been carried out within the frame of the National Centre for Competence in Research' PlanetS' supported by the Swiss National Science Foundation (SNSF). The authors acknowledge the financial support of the SNSF. The research leading to these results has received funding from the European Research Council (ERC) under the European Union's Horizon 2020 research and innovation programme (grant agreement No. 679633: Exo-Atmos). This work was supported by FCT - Fundação para a Ciência e a Tecnologia through national funds and by FEDER through COMPETE2020 - Programa Operacional Competitividade e Internacionalização by these grants: UID/FIS/04434/2019; UIDB/04434/2020; UIDP/04434/2020; PTDC/FIS-AST/32113/2017 $\&$ POCI-01-0145-FEDER-032113; PTDC/FIS-AST/28953/2017 $\&$ POCI-01-0145-FEDER-028953; PTDC/FIS-AST/28987/2017 $\&$ POCI-01-0145-FEDER-028987 $\&$ PTDC/FIS-OUT/29048/2017 $\&$ IF/00852/2015.. This project has received funding from the European Research Council (ERC) under the European Union's Horizon 2020 research and innovation programme (project Four Aces grant agreement No 724427). O.D.S.D. is supported by by national funds through Fundação para a Ciência e Tecnologia (FCT)
in the form of a work contract (DL 57/2016/CP1364/CT0004) and project related funds (EPIC: PTDC/FIS-AST/28953/2017 $\&$ POCI-01-0145-FEDER-028953). The INAF authors acknowledge financial support of the Italian Ministry of Education, University, and Research with PRIN 201278X4FL and the "Progetti Premiali" funding scheme. V.A. acknowledges the support from FCT through Investigador FCT contract nr. IF/00650/2015/CP1273/CT0001. This work has made use of data from the European Space Agency (ESA) mission {\it Gaia} (\url{https://www.cosmos.esa.int/gaia}), processed by the {\it Gaia} Data Processing and Analysis Consortium (DPAC, \url{https://www.cosmos.esa.int/web/gaia/dpac/consortium}). Funding for the DPAC has been provided by national institutions, in particular the institutions participating in the {\it Gaia} Multilateral Agreement.
\end{acknowledgements}
\bibliographystyle{aa}
\bibliography{bib}

\begin{appendix}
\section{Binary masks or template matching for CCFs}\label{AppendixA}
In this Appendix, we describe the main difference between a CCF obtained through cross-correlation with a model or with a binary mask for atmospheric characterization. Both methods rely on different assumptions that are discussed below, except that they both need a line-list.  The line-list selection is discussed in Appendix\, \ref{AppendixB}. Nonetheless, the main idea is similar and consists of co-adding hundreds of lines to create an average line. Molecular lines in a planetary spectrum have different depths, and when hundreds of them are combined, the weakest lines will decrease the SNR. Therefore, it is important to give proper weight to each line to maximize the SNR.
\subsection{Template matching} \label{AppA_Temp}
The template matching is the most commonly used technique to detect molecular species in exoplanet atmospheres \citep[e.g.][]{snellen_orbital_2010,birkby_detection_2013,brogi_rotation_2016}. It consists in building an atmospheric model that depends on the line positions, line cross-sections, temperature, mean-molecular weight, molecule abundance, cloud deck, and haze scattering. The model produced has thus a proper weight of the intrinsic depth of the lines at each pixel. However, this technique requires that we have prior knowledge of the atmospheric properties or that we guess them well. Hence, a grid of models is required, which is computationally heavy. Recently, \cite{brogi_retrieving_2019,gibson_detection_2020} have proposed and tested bayesian frameworks to find the best model maximizing the S/N. These frameworks are able to retrieve abundances, reconstruct the temperature-pressure profile and can be used on low- and high-resolution datasets simultaneously. Finally, this method does not allow to easily retrieve basic properties (it requires the new bayesian framework) of the average line such as its amplitude, its FWHM, or the continuum dispersion that can be useful to compare to other results directly. 
\subsection{Binary masks}
This method for atmospheric studies has barely been used so far \citep{allart_search_2017,pino_diagnosing_2018,pino_neutral_2020}. It consists of building a binary mask (with an aperture width of one pixel) and summing the transmitted flux at each wavelength. The mask apertures are centered on the theoretical line positions that depend only on the temperature and that are associated with a given weight. Accordingly, this technique relies on three parameters: line positions, temperature, and weights, and is not defined over the full spectral range. The major and more problematic questions linked to this method is how the weights have to be set. At the difference with stellar masks used to extract radial velocities, it is not possible to measure the line contrasts for a given planet and apply it to all exoplanets. The weights of exoplanet masks encompass all the other parameters described in appendix\,\ref{AppA_Temp}. \cite{hoeijmakers_spectral_2019} used a mixed approach between template matching and binary mask, which consists in using designed templates, for a given planet, to set the weights. Here, we fixed the weight to one for each line to minimize model assumptions and follow Occam's razor principle. Now the difficulty lies on how many lines have to be included in the mask starting from the strongest to maximize the S/N. On the one hand, including only a handful of lines implies that the CCF signal is high, as well as the noise. On the other hand, including all the lines minimizes the noise but at the expense of removing the signal due to the thousands of weakest lines. To maximize the S/R, it is thus necessary to optimize the number of lines in the mask, but this number cannot be known a priori. While the template-matching needs several parameters, the here-proposed technique relies only on two parameters (temperature and number of lines) that can be explored relatively quickly. Moreover, the produced average line can be expressed on its basic properties.

\subsection{Complementary between both methods}
Nevertheless, the two methods are not opposite and are, in fact, complementary. This can be seen as a two-steps approach. First, the binary mask method is applied, as it is faster, to retrieve the basic properties of the average line and give a first constraint on the temperature range. Then, the template matching method comes in a second step to retrieve in greater detail the atmospheric properties (e.g., temperature-pressure profile, abundances, cloud deck). We also note that in \cite{pino_neutral_2020}, the authors proposed a binary masks bayesian framework, benchmarked it with template matching bayesian framework \citep{brogi_retrieving_2019} and retrieved detailed information on the atmosphere of KELT-9b.

\section{Line-list comparison for water in the visible}\label{AppendixB}
\begin{figure}[h]
\resizebox{\hsize}{!}{\includegraphics{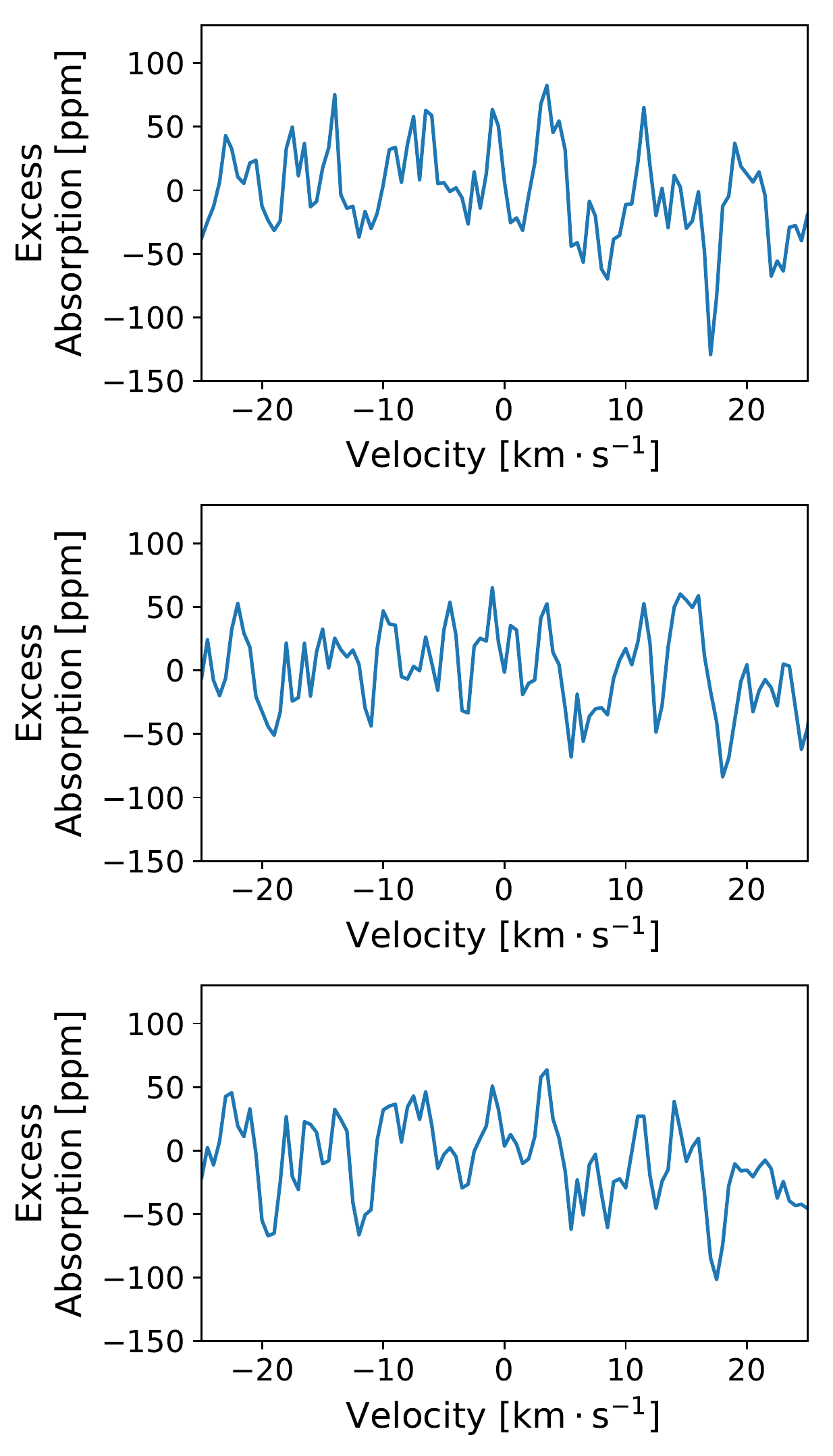}}
\caption[Line_list]{CCFs for the ExoMol BT2 (\textit{top panel}), the ExoMol Pokazatel (\textit{middle panel}) and the HITEMP2010 (\textit{bottom panel}) with a water binary mask at 1400\,K and 1600 lines.}
\label{Line_list}
\end{figure}

Various molecular line-lists (e.g., HITEMP, ExoMol, ...) can reproduce broad-band detection like the 1.3\,microns water band. However, when going toward higher resolutions where individual lines are resolved ($\mathcal{R}\sim$100000), the line positions and cross-sections are less reliable\footnote{There is a strong effort put into this such as the ExoMol and HITEMP projects}. \citep{gandhi_molecular_2020} studied the impact of different line-lists in the infrared. The authors showed that water line-lists mostly diverged for temperature above 1200\,K and for low opacity lines. On the contrary, the strongest water lines are in agreement between line-lists in term of position and intensity. Here, we focus on the search for water vapor in the visible range (7000 to 7500\,\AA) and we compare three existing line-lists: ExoMol BT2 \citep{barber_high-accuracy_2006}, Exomol Pokazatel \citep{polyansky_exomol_2018} and HITEMP2010 \citep{rothman_hitemp_2010}. We applied the same technique as described in Sect\,\ref{Sub_sec_CCF} for each line-list. Fig.\,\ref{Line_list} shows for each line-list, the CCFs for a temperature of 1400\,K, and 1600 lines. It clearly reveals that the three line-lists present the same broad patterns, but none of them shows a water signature. There are two possible explanations for this: either the water signal is muted by the cloud-deck, as we proposed, or all line-lists are equally bad in the visible and thus prevent the search for water vapor.

\section{ Injection model and constraints from binary masks}\label{AppendixC}
\begin{figure}[h]
\centering
\includegraphics[width=\columnwidth]{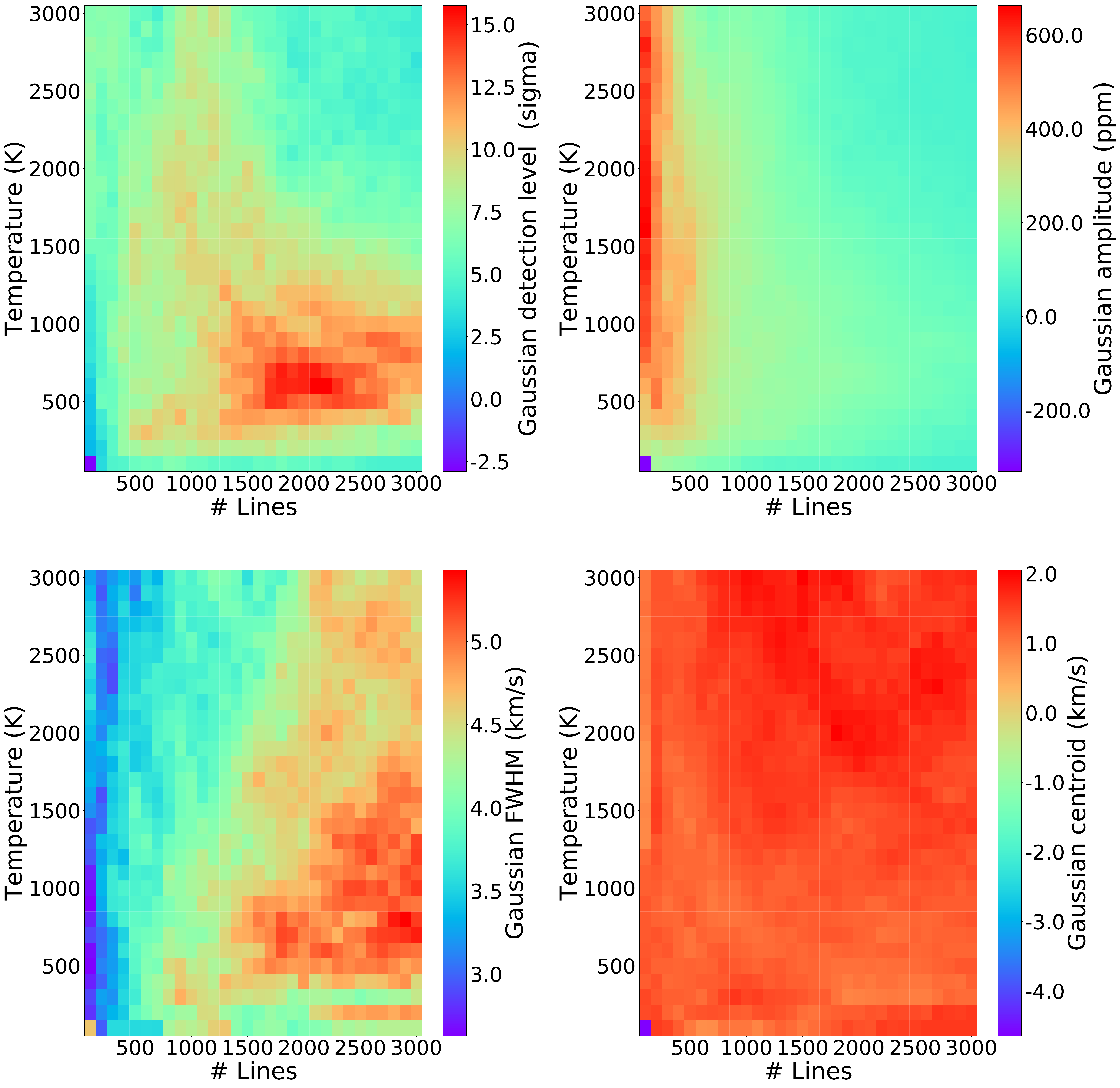}
\caption[Injection_test_map]{Temperature vs. number lines map representing the Gaussian fit outputs for each CCF set, from \textit{top} to \textit{bottom}: the detection level, the amplitude, the FWHM and the centroid position. The detection map shows clearly that the binary mask approach can constrain the temperature of the atmosphere and the number of lines (i.e., the weights) required. The amplitude map behaves as expected with the higher signal when fewer lines are considered, but they are the strongest. Conclusions from the Gaussian FWHM and centroid are more difficult to draw, but at least where the detection level is highest, the FWHM and the centroids vary smoothly as a function of temperature and number of lines. This can be used to give more confidence in the detection.}
\label{CCF_map_T_line_injected}
\end{figure}

\begin{figure}[h]
\includegraphics[width=\columnwidth]{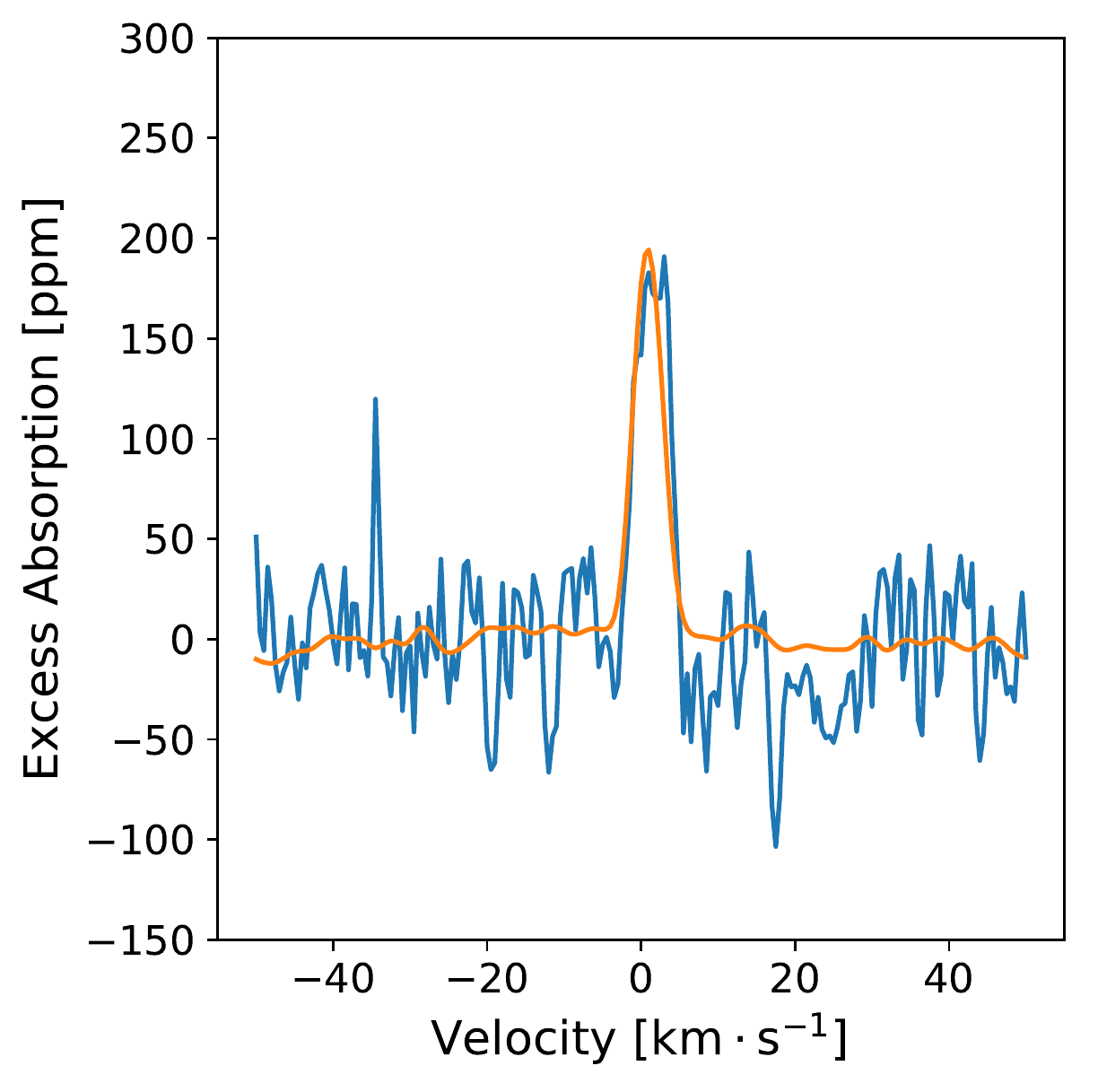}
\caption[Injection_test_best_fit]{Measured CCF with the injected model with the water vapor mask at 1400\,K and 1600\,lines in blue. The same mask is used to compute the modeled CCF shown in orange. This mask is used as it maximizes the detection level with the temperature fixed to T$_{eq}$.}
\label{CCF_map_T_line_injected_best}
\end{figure}

In this Appendix, we injected the cloud-free model described in Sect\,\ref{model_description} to see if our reduction steps described in Sect\,\ref{Sec_method} could modify or erase a planetary signature. Our second goal is to apply our CCF binary mask framework to the ESPRESSO data when a planetary signature is present. To do so, we injected the cloud-free model on each in-transit spectrum before applying the telluric correction. We then reran the \texttt{Molecfit} correction and the \texttt{RASSINE} normalization before extracting the transmission spectrum and deriving the CCFs.\\
Fig.\,\ref{CCF_map_T_line_injected} presents the temperature vs. number lines map described in Sect.\,\ref{Sub_sec_CCF} and reports the best value of each parameters of the Gaussian fit applied to each CCF. First, the detection map (top row) shows clearly that the injected water vapor signature is detected and that the binary mask procedure allows to constrain the water vapor temperature and additionally, the weights (expressed in the number of strongest lines) of the lines. Second, while the Gaussian amplitude varies as expected (strongest amplitudes when fewer lines are included in the mask), it does not provide a strong constraint on the signal apart from an indicative mean amplitude of the water lines. Finally, the Gaussian FWHM and centroid are used to verify that the detection is not obtained for low or high-frequency noise at a velocity far from 0\,km$\cdot$s$^{-1}$.\\
Fig.\,\ref{CCF_map_T_line_injected_best} shows the observed CCF that maximizes the detection for the equilibrium temperature, and the model CCF using the same mask. One can see that the injected signal is well detected, and it is also recovered in the 1-$\sigma$ uncertainty. Therefore, our reduction steps, including telluric correction, normalization, wiggles removal, interpolation due to Doppler-shifts, do not modify or erase a planetary signature.\\
As a general comment, the decision to look at the CCFs at T$_{eq}$ and not at the \textit{HST}/WFC3 retrieval temperature used to compute the model can be explained in two points. Firstly, we want to follow a model-independent approach as much as possible, which favors choosing the equilibrium temperature and not the one from a retrieval. Secondly, \cite{macdonald_why_2020} discusses in great detail why \textit{HST}/WFC3 retrievals applied to the water band at 1.3\,microns found temperatures lower than the equilibrium temperatures. The explanation is based on the lack of sophistication of those models that are generally 1-D and do not allow different compositions between terminators, thus leading to erroneous abundance, temperature, and temperature gradient.

\section{Radial velocities}\label{AppendixD}
In this appendix, we show the radial velocities of WASP-127b phase-folded around mid-transit to illustrate the classical Rossiter-Mclaughlin effect.

\begin{figure}[h]
\includegraphics[width=\columnwidth]{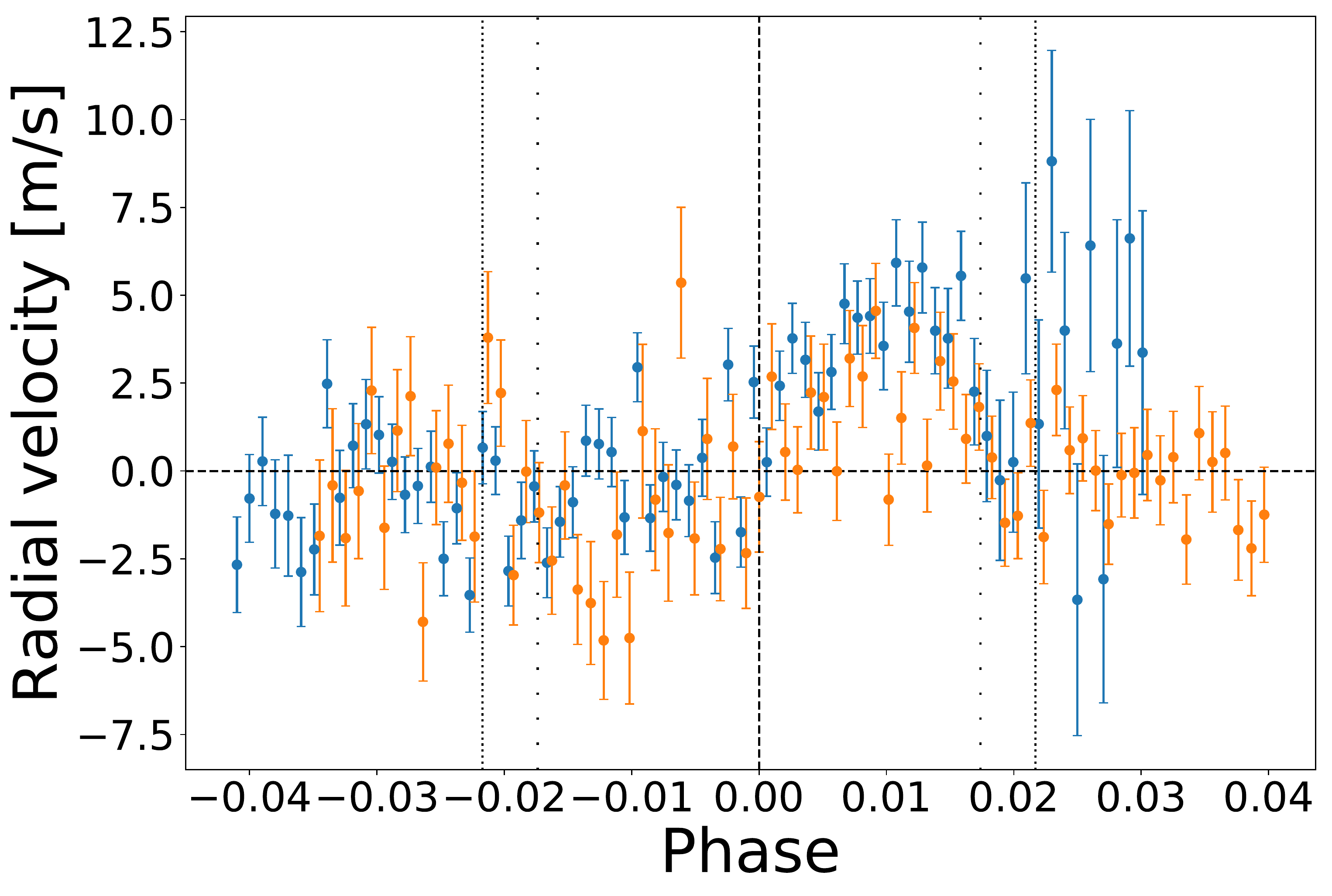}
\caption[RVs]{Radial velocities of WASP-127b for both transits (respectively blue and orange for the 2019-02-24 and 2019-03-17) phased-folded. The radial velocities have been corrected from the v$_{syst}$ and the keplerian model. We can clearly see the RM effect expected for a retrograde orbit with negative velocities during the first part of the transit followed by positive radial velocities. The vertical dashed line is the mid-transit while the dense vertical dotted lines correspond to t$_1$ and t$_4$ and the vertical dotted lines are t$_2$ and t$_3$. The horizontal dashed line corresponds to null velocity.}
\label{RVs}
\end{figure}

\section{Impact of the normalization}\label{AppendixE}
In this appendix, we show the impact of the \texttt{RASSINE} \citep{cretignier_rassine_2020} normalization on the transmission spectrum. The wiggles are well suppressed by the \texttt{RASSINE} normalization but few residuals are still present. This is due to the choice of the preliminary mode of spectra time-series normalization of \texttt{RASSINE} rather that the complete version since this latter could absorb true planetary signatures in the Savitchy-Golay filtering. This was discussed by the authors in the Fig.\,4 of their paper.

\begin{figure}[h]
\includegraphics[width=\columnwidth]{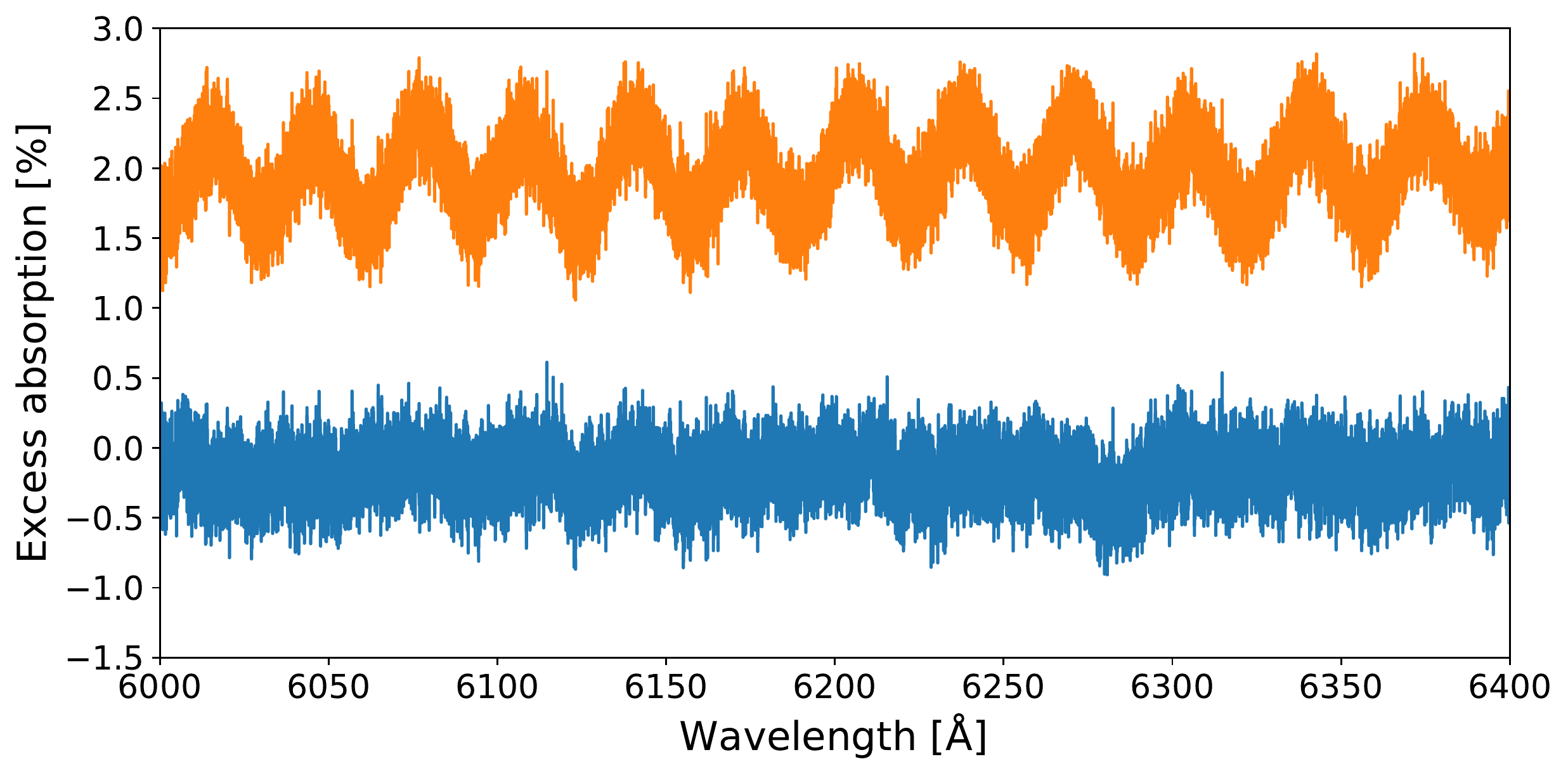}
\caption[Norm_impact]{Transmission spectrum with (blue) or without (orange) the  \texttt{RASSINE} normalization for the 2019-02-24 transit.}
\label{Norm_impact}
\end{figure}

\end{appendix}
\end{document}